\documentclass[%
 aip,
 amsmath,amssymb,
 reprint,%
]{revtex4-1}

\usepackage{graphicx}
\usepackage{dcolumn}
\usepackage{bm}

\usepackage[utf8]{inputenc}
\usepackage[T1]{fontenc}
\usepackage{mathptmx}

\begin{document}

\preprint{AIP/123-QED}

\title[A project of advanced solid-state neutron polarizer for PF1B]{A project of advanced solid-state neutron polarizer for PF1B instrument at ILL}

\author{A.K. Petukhov}
 \email{petukhov@ill.eu.}
 
\author{V.V. Nesvizhevsky}
 \email{nesvizhevsky@ill.eu.}

\author{T. Bigault}

\author{P. Courtois}

\author{D. Jullien}

\author{T. Soldner}

 \affiliation{Institut Max von Laue -- Paul Langevin, 71 avenue des Martyrs, Grenoble, France, F-38042}

\date{\today}

\begin{abstract}
Among polarizers based on the neutron reflection from Super-Mirrors (SM), solid-state neutron-optical devices have many advantages. The most relevant is 5-10 times smaller size along the neutron beam direction compared to more traditional air-gap devices allowing to apply stronger magnetic fields to magnetize SM. An important condition for a good SM polarizer is the matching of the substrate SLD (Scattering Length Density) with the SM coating SLD for the spin-down component. For traditional Fe/Si SM on Si substrate, this SLD step is positive when a neutron goes from the substrate to the SM, which leads to a significant degradation of the polarizer performance in the small Q region. This can be solved by replacing \cite{petukh16} single-crystal Si substrates by single-crystal Sapphire/Quartz substrates. The latter show a negative SLD step for the spin-down neutron polarization component at the interface with Fe and, therefore, avoid the total reflection regime in the small Q region. In order to optimize the polarizer performance, we formulate the concept of Sapphire V-bender. We perform ray-tracing simulations of Sapphire V-bender, compare results with those for traditional C-bender on Si, and study experimentally V-bender prototypes with different substrates. Our results show that the choice of substrate material, polarizer geometry as well the strength and quality of magnetizing field have dramatic effect on the polarizer performance. In particular, we compare the performance of polarizer for the applied magnetic field strength of $50 mT$ and $300 mT$. Only the large field strength ($300 mT$) provides an excellent agreement between the simulated and measured polarization values. For the double-collision configuration, a record polarization $>0.999$ was obtained in the neutron wavelength band of $0.3-1.2 nm$ with only $1\%$ decrease at $2 nm$. Without any collimation, the polarization averaged over the full outgoing capture spectrum, $0.997$, was found to be equal to the value obtained previously only using a double polarizer in the "crossed" (X-SM) geometry \cite{kreuz05}. These results are applied in a full-scale polarizer for the PF1B instrument.
\end{abstract}

\maketitle

\section{\label{sec:level1}Introduction}

PF1B is a user facility at the Institut Laue-Langevin (ILL) in Grenoble, France, for experiments in elementary particle and nuclear physics using polarized or unpolarized cold neutrons \cite{abele06}. An important component of PF1B is a device that produces a neutron beam of a large area with a high polarization in the broad range of neutron wavelengths $\lambda =0.2-2.0 nm$. To our knowledge, with present technology a reflection-type SM-based polarizer is the only type of device which can fulfill such requirements.

For carrying out experiments of different type, PF1B has to provide several options including: a maximum total flux of polarized neutrons over a large beam cross section \cite{vesna08,vesna17}, a maximum flux density of polarized neutrons over a relatively small beam cross section \cite{goen07,gagar16}, and ultra-high precision of the knowledge of the polarized neutron beam properties \cite{schum07}. While in the first two cases the mean polarization can be moderate (typically $\overline{P_n}>98\%$), the latter option requires ultra-high polarization ($\overline{P_n}>99.7\%$) to minimize the systematic uncertainty.

The existing PF1B polarizer was built using the traditional technology of polarizing benders \cite{mezei77,drab77,schaer86,majk95,maru07,krist08,mezei89}: 30 channels $80 cm$ long, air gaps of $2 mm$, borated float glass substrates $0.7 mm$ thick, $m=2.8$ SM coatings \cite{schaer86,and94,cour13}, $80x80 mm^2$ cross-section, $300 m$ radius, $120 mT$ applied magnetic field \cite{sol02}. This polarizer was manufactured, the coating being performed at ILL, and installed in 2004 downstream the ballistic neutron guide H113 \cite{abele06} with the effective neutron capture flux of $2 \cdot 10^{10} n/cm^2/s$. Its measured transmission was $\sim 50\%$ for the "good" polarization component, and the mean polarization was $\sim 98.5\%$. When ultra-high polarization was required, we installed a second similar polarizer in the "crossed geometry", thus obtaining the mean polarization of $\overline{P_n}= 99.7\%$ and the transmission $\sim 25\%$ \cite{kreuz05} for the "good" polarization component. During 15 years of successful exploitation, this polarizer was irradiated with a very high neutron fluence, which resulted in significant radiation damage to the mirrors' glass substrates (mainly by charged particles products from the reaction $^{10}B(n,\alpha)$ in the glass substrate). It is also strongly activated, mainly due to the presence of Co in the SM coatings, which makes its handling more complicated. In this paper, we present a project of a new advanced polarizer for PF1B with an improved polarization and transmission performance and free from radiation damage and activation issues.

\section{\label{sec:level1}Polarizer design}

To overcome the main drawbacks of the existing polarizer: high activation of Co in the Co/Ti SM coating and its neutron induced degradation caused by the $^{10}B(n,\alpha)$ reaction in glass substrates, - we chose to build a solid-state polarizer with a Fe/Si coating \cite{majk95,krist98,hog99,stu06,big09,wild11}. Other immediate advantages of a solid-state polarizer are compactness, thus allowing a magnetic system with better performance, and a more favorable ratio of channel to inter-channel width. In that version, the polarizer is traditionally made of stacks of thin ($150-200 \mu m$ thick) single crystal Si wafers, coated on both sides with Fe/Si SM coatings terminated by Gd absorbing layers. Each Si plate confined between two reflecting SMs serves as a spin-dependent neutron guide for neutrons entering through the entrance edge of the plate. 

In an ideal polarizer of this type, one neutron spin-component penetrates through the SM coating without reflection and then is absorbed in the capping Gd layer. The other, "good" spin-component is reflected by the SM coating and propagates to the exit edge of the Si substrate. Such neutron behaviour requires perfect matching of the Scattering Length Densities (SLD) of SM materials for one neutron spin component and a high contrast for the other (SLD: ${\rho}^{\pm}=\sum_{j}N_{j}b_{j\pm}$, where $N_{j}$ is the number density of nuclei and $b_{j\pm}$ is the spin-dependent neutron scattering length for element/isotope $j$). In practice, materials used for SMs never match perfectly. Materials of interest for neutron polarization applications are listed in Table \ref{Table 1} \cite{sears92}. In order to optimize the SLD matching for the "-" spin component, some parameters are tuned during the coating process, resulting in our case in an increase of the SLD for the Si layers, making it closer to $\rho^{-}_{Fe}$ \cite{hog99}. 

\begin{table}
\caption{\label{Table 1} Neutron scattering length density [$10^{-6}$\AA$^{-2}$] of various bender materials. Values from \cite{sears92}.}
\begin{ruledtabular}
\begin{tabular}{ccccccccc}
&Fe&\mbox{Co}&\mbox{Ni}&\mbox{Ti}&\mbox{Si}&\mbox{SiO2}&\mbox{Al2O3}&\mbox{Gd}\\
&&\mbox{}&\mbox{}&\mbox{}&\mbox{}&\mbox{quartz}&\mbox{sapphire}&\mbox{}\\
\hline
$\rho$&8.02&\mbox{2.27}&\mbox{9.4}&\mbox{-1.93}&\mbox{2.08}&\mbox{4.19}&\mbox{5.72}&\mbox{2.24-i0.325}\\
$\rho^{+}$&12.97&\mbox{6.38}&\mbox{10.86}&\mbox{}&\mbox{}&\mbox{}&\mbox{}&\mbox{}\\
$\rho^{-}$&3.08&\mbox{-2.14}&\mbox{7.94}&\mbox{}&\mbox{}&\mbox{}&\mbox{}&\mbox{}\\
\end{tabular}
\end{ruledtabular}
\end{table}
For neutrons incident on the Fe/Si coating from inside the Si substrate, the spin-up reflectivity is governed by the interference pattern of all layers in the SM while the spin-down reflectivity is mostly governed by the SLD step from Si to the thickest layer of Fe:
\begin{eqnarray}
\Delta\rho^{-}=\rho^{-}_{Fe}-\rho_{Si} 
\label{SLDstep}.
\end{eqnarray}
For solid-state Fe/Si SM on Si substrate, this SLD step is positive leading to the total reflection of neutrons incident at small grazing angles:
\begin{eqnarray}
\Theta^{-}_c=\lambda\sqrt{\Delta\rho^{-}/\pi}
\label{Teta},
\end{eqnarray}
where $\lambda$ is the neutron wavelength in \AA. For sufficiently small grazing angles $\Theta<\Theta^{-}_c$, both spin-components are totally reflected by the interface thus resulting in zero polarizing efficiency.

\begin{figure}
\includegraphics[width=\linewidth]{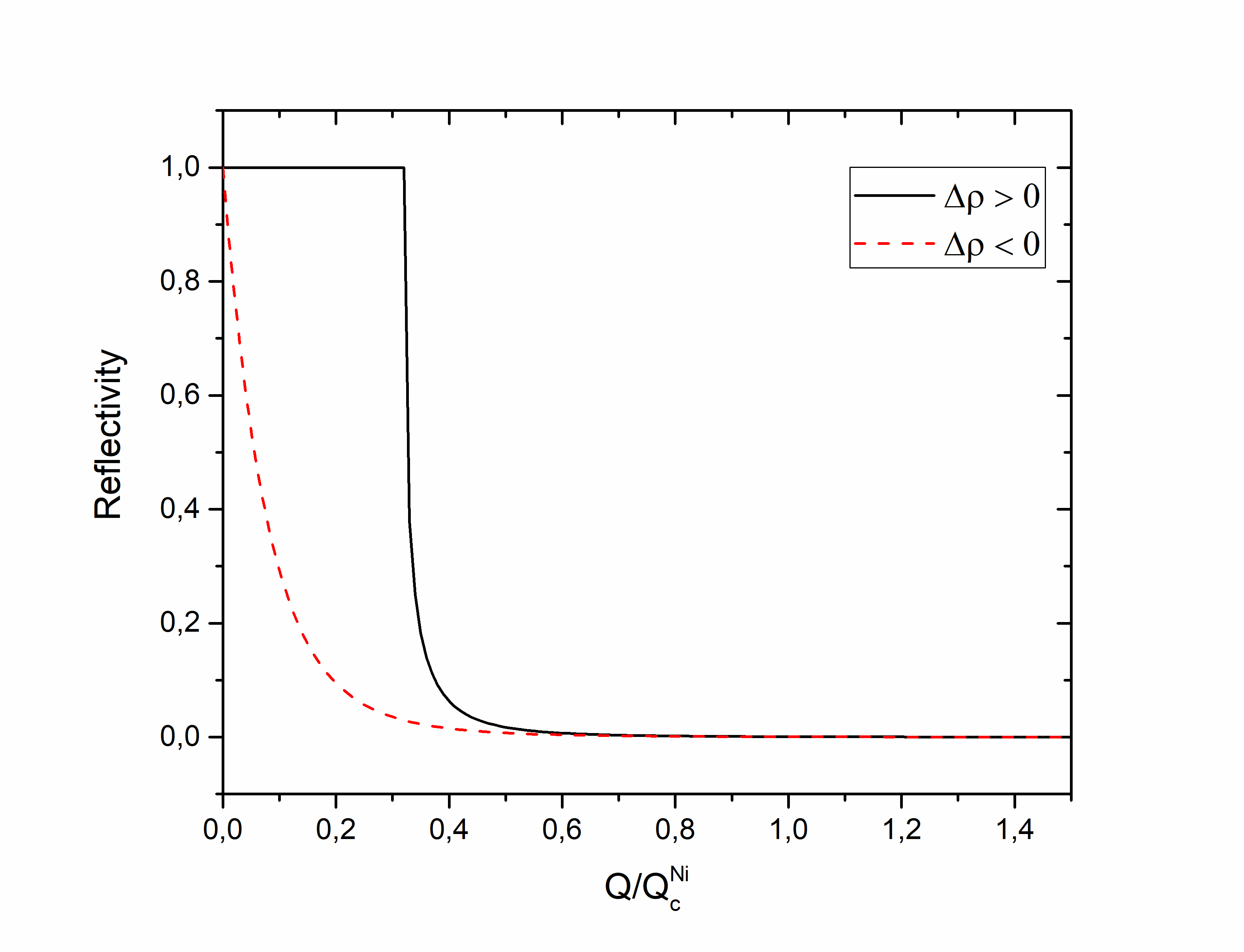}
\caption{\label{reflectivity} Calculated reflectivity curves as a function of momentum transfer ($Q^{Ni}_c$ is the critical momentum transfer for natural Ni) for spin-down neutrons incident from substrate on a thick layer of Fe for positive and negative steps in SLD. The amplitude of the step is the same $\rho=10^{-6}$\AA$^{-2}$.}
\end{figure}

In our previous paper \cite{petukh16}, we showed that the replacement of single-crystal Si substrates by single-crystal Quartz or Sapphire wafers results in a negative SLD step for spin-down neutrons that eliminates the total reflection regime, see Fig. \ref{reflectivity}, and thus expands the polarizer bandwidth into the low Q region. 

The modern progress in production of high-quality single-crystal Quartz and Sapphire wafers makes them readily available for our applications with prices close to those for Si wafers. In particular, surface polishing down to a roughness of about $0.5 nm$ (r.m.s.) are now available, thus ensuring sufficient SM performance.

As it follows from Table \ref{Table 1}, thin single-crystal Quartz wafers are expected to be the first choice for substrates for Fe/Si solid-state polarizers. Indeed, for spin-down neutrons passing from a Quartz substrate to a thick Fe layer, the SLD step is negative and relatively small ($\rho^{-}_{Fe}-\rho_{Quartz}=-1.1\cdot10^{-6}$ [\AA$^{-2}$], $\rho^{-}_{Fe}-\rho_{Sapphire}=-2.6\cdot10^{-6}$ [\AA$^{-2}$]). However, we found that single-crystal Quartz wafers are very fragile. During manipulations, the sharp edges of quartz plates tend to crumble and produce a large amount of quartz dust particles with a typical size of the order of $1-10 \mu m$. These fragments are randomly located on the surface of the plates and significantly complicate the assembly of the plates parallel to each other thus leading to a significant angular spread of the reflected neutrons.

We also found that Sapphire plates are much more rigid than both Quartz and Si plates, and their edges do not crumble into such dust-like fragments in SM deposition process. Moreover, residual stress in he SM coating, induced by the SM film growth process, results in significant bending of such thin substrates ($0.18 mm$) \cite{sto09}. In the case of Sapphire, due to the higher elastic constants, this effect is less pronounced for the same stress value, therefore the substrate deformation after coating is reduced. Taking into account these practical aspects, we have decided to produce a solid-state Si/Fe polarizer for PF1B using single-crystal Sapphire substrates. 

Usually, the stack of such substrates (with thickness \textit{d} and length \textit{L}) is bent elastically (with radius \textit{R}) to avoid neutron trajectories without collisions with the SM coatings. Thus, the bending angle $\gamma_c$ has to meet the condition   
\begin{eqnarray}
\gamma_c>8d/L.
\label{bending angle}
\end{eqnarray}
This technique works well if the number of plates is limited to a few dozen. 

However, the required cross section of a PF1B polarizer is as large as at least $80x80 mm^2$. For the thickness of Sapphire plates of $0.18-0.20 mm$, the number of plates in the stack would be as large as 400-500. Uniform bending of such a stack is challenging \cite{majk95}.

\begin{figure}
\includegraphics[width=\linewidth]{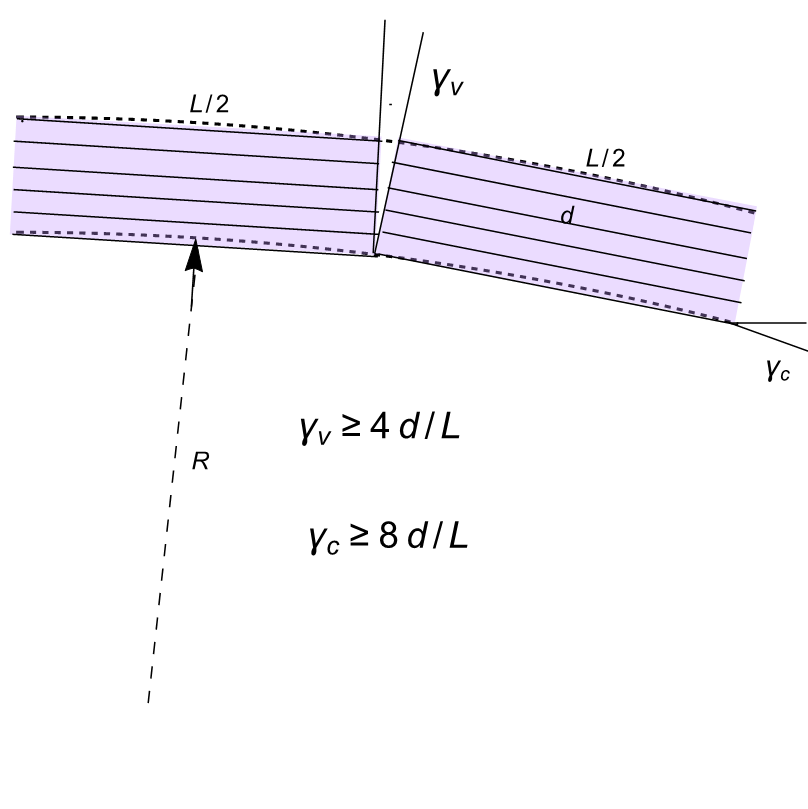}
\caption{\label{V} The V-bender geometry: two stacks of both sides SM-coated plane-parallel substrate plates (with thickness \textit{d} and length \textit{L}/2) are tilted by angle $\gamma_v$ with respect to each other. To avoid the "direct view", the angle has to meet condition (\ref{tilte angle}). The C-bender geometry is shown (dashed lines) for comparison.}
\end{figure}

An alternative solution is to use polygonal geometry. In the simplest case, the continuous bend is replaced by two stacks tilted by the angle $\gamma_v$ with respect to each other, see Fig. \ref{V}.
\begin{eqnarray}
\gamma_v>4d/L.
\label{tilte angle}
\end{eqnarray}

\begin{figure}
\includegraphics[width=\linewidth]{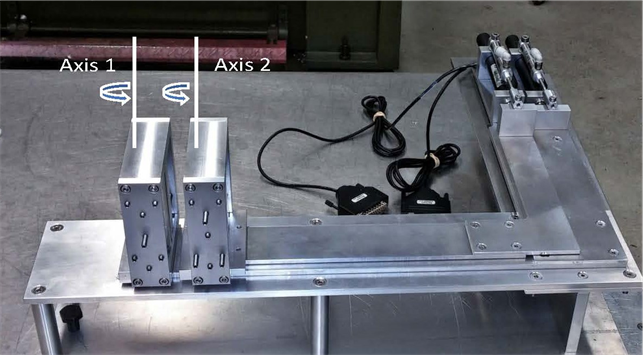}
\caption{\label{prototype} A prototype of the adaptable V-bender.}
\end{figure}

The V-bender geometry greatly simplifies the assembling of the stacks. Each of the two stacks can be produced independently. Each stack can be individually motorized, thus converting the V-bender into an adaptable optics device. This option makes it possible to remotely control the main characteristics of the polarizer: polarization $P$, transmission $T$ and critical wavelength $\lambda_c$ in accordance with the experimental requirements. Fig. \ref{prototype} shows a prototype of the adaptable V-bender.

\begin{figure}
\includegraphics[width=\linewidth]{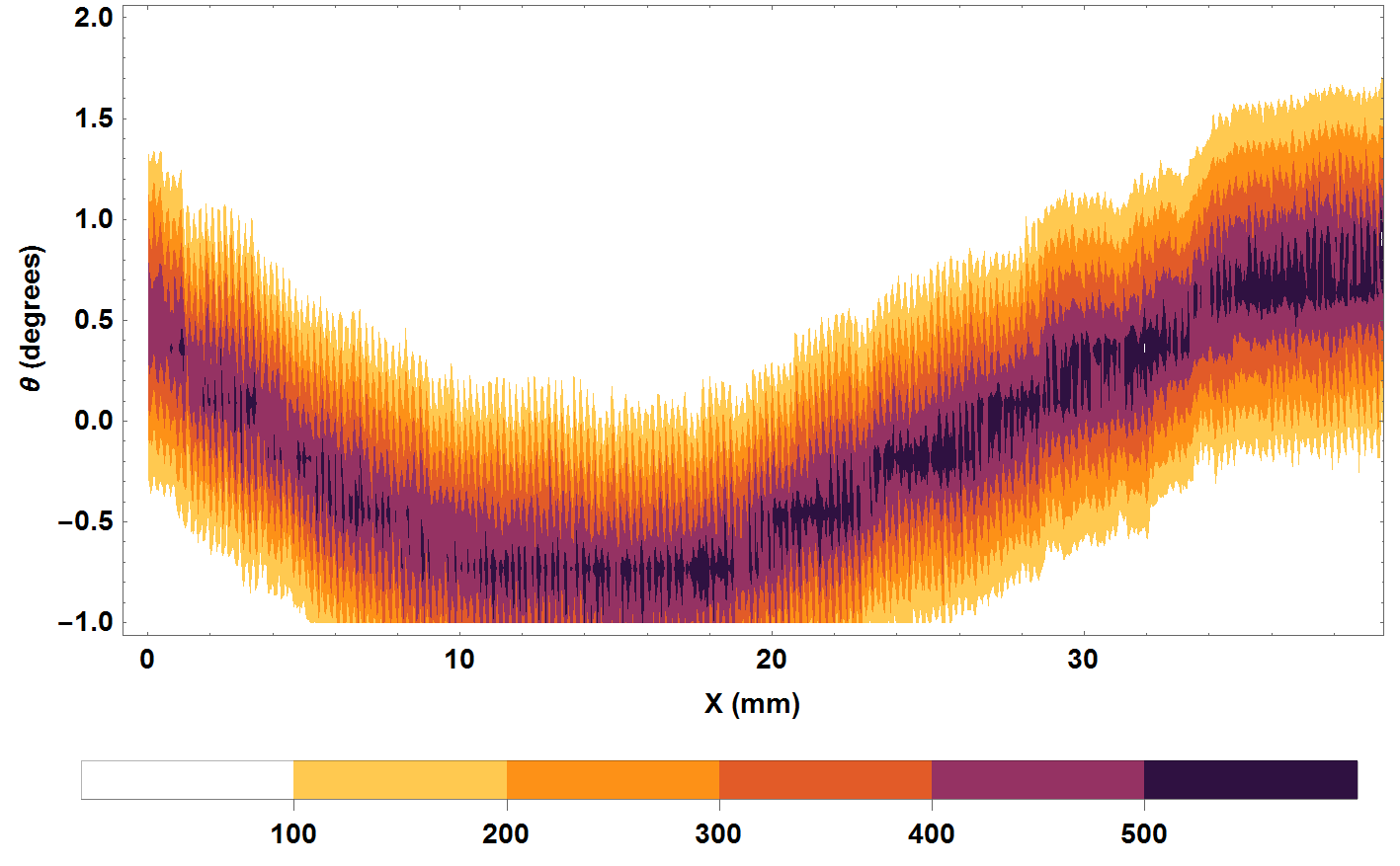}
\caption{\label{random walks} Counts of neutrons (illustrated with different colors as shown on the scale below the figure) transmitted through a solid-state Soller collimator (the acceptance profile) measured at T3 instrument at ILL as a function of the position in the collimator stack (X) and angle of the collimated incident beam ($\Theta$). The collimator consists of a stack of 200 Gd-coated single-crystal Si substrates with a thickness of $200 \mu m$ assembled without control of the mutual plate inclination angle during the collimator assembling. For the perfect plane-parallel geometry of the plates and their perfect assembling, all the acceptance profiles have to be aligned along a straight line.}
\end{figure}

The simple geometry of each stack allows optical in-situ control of the inclination angle of each coated plate in the package. Imperfections of the plane-parallel geometry of individual substrates, as well as dust particles falling between them, would result in a scatter of the tilt angles. Errors in setting the tilt angle, and hence the angle of reflection of neutrons, would accumulate with an increase in the number of substrates in the stack according to the law of "random walks", see Fig. \ref{random walks}.

The primary effect of this angular spread is the degradation of the polarizer transmission for the "good" spin-component, provided that it is comparable to the critical angle of the polarizer (typically $10-20 mrad$). This mechanism may explain the discrepancy between the expected and measured transmission probabilities often observed in experiments with solid-state neutron-optical devices \cite{shah14}. Some polarization loss may also be attributed to this effect, since this angular spread may create "wrong spin leaks", by making some neutron trajectories without collision possible.

An important feature of the V-Bender is the absence of pronounced Bragg dips in the transmission. Such dips were observed in experiments with solid-state S- and C-benders when the angular divergence of the neutron beam is comparable with the angle of continuous bending $\gamma_c$ \cite{stun06,shah14}. For reflection of neutrons from a flat perfect crystal, the acceptance angle of Bragg reflection is very small (typically $1-10 \mu rad$) and the corresponding dip is completely washed out by the angular divergence of the incident beam (typically a few tens $mrad$).

Though some V-bender geometries were used rather early as neutron polarizers \cite{mol63}, it seems that since the advent of SM more interest has been put in C-bent or S-bent devices, without always considering the various advantages of V-bender geometry.

\section{\label{sec:level1}Polarizer simulations}

In this section, we compare Monte-Carlo ray-tracing simulation predictions for sapphire V-Bender and for traditional Si C-Bender. To cover the wavelength band of $0.3-2.0 nm$, we use the same “inverse” scheme $m = 3.2$ Fe/SiN$_x$ SM coating, which was previously used for the production of solid-state S-Bender \cite{stun06}. The term "inverse" refers to the sequence of deposition of a SM coating, starting with a thicker layer as opposed to the sequence in SMs used for neutrons incident from air. Note that with such a sequence in a solid-state polarizer, the very first layer visible by neutrons is the thickest as in the case of air-gaped devices. An absorbing Gd layer also has to be coated on top of the SM, so that neutrons which are transmitted through the SM do not get out of the polarizer. In order to guarantee that these neutrons are absorbed rather than reflected, even at low Q values, an anti-reflecting and absorbing Si/Gd multilayer based on the same principle as in \cite{schar94} was designed, replacing the simple Gd layer. 

\begin{figure}
\includegraphics[width=1.15\linewidth]{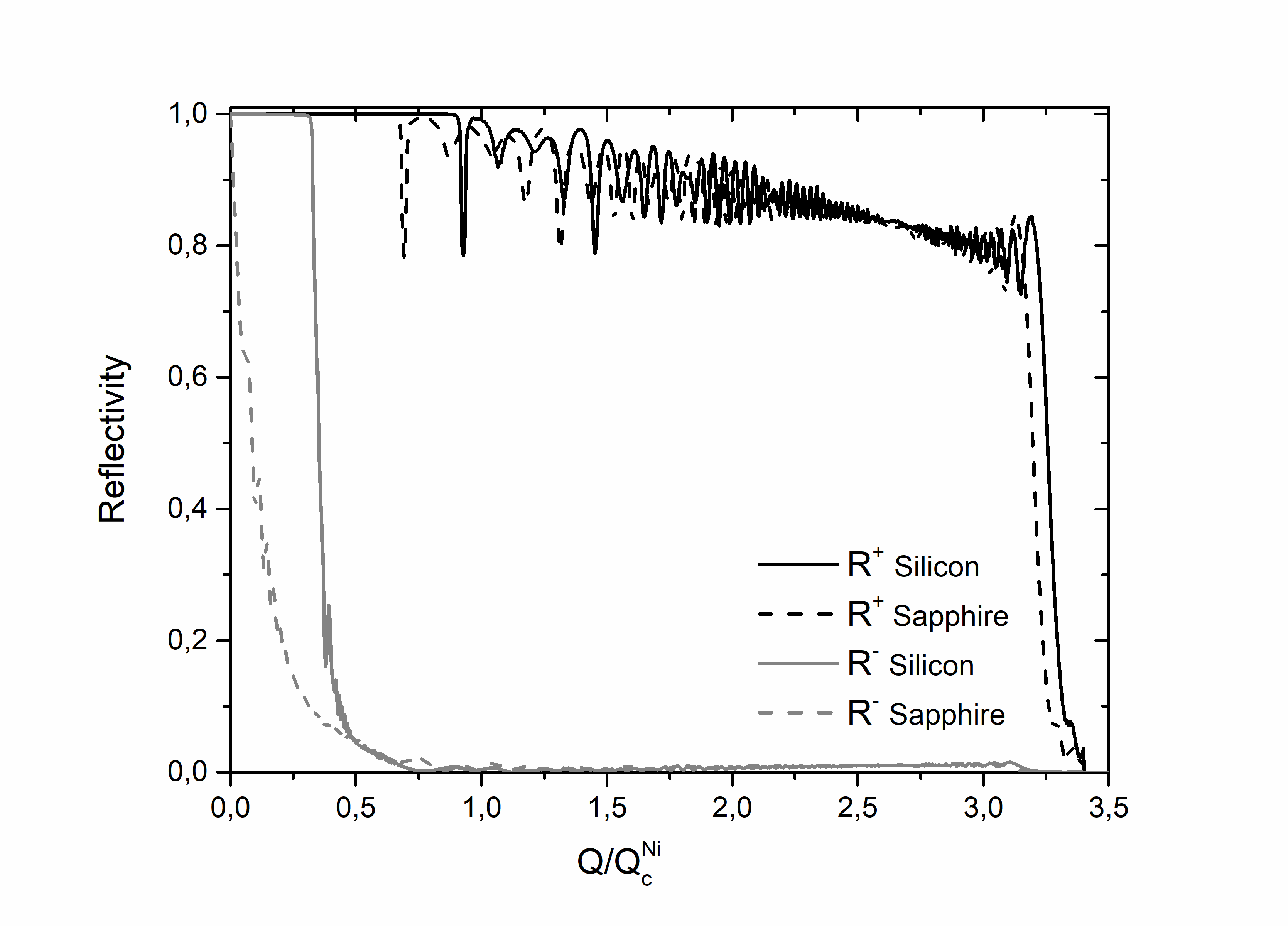}
\caption{\label{SiSaReflectivity} Curves of simulated spin-dependent reflectivity ($R^{+}$ - black lines, $R^{-}$ - grey lines) for $Fe/SiN_x$ SM deposited on different substrate materials. Neutrons are incident on the interface from the substrate.}
\end{figure}

Fig. \ref{SiSaReflectivity} depicts the spin-dependent reflectivity for neutrons incident on the SM interface from inside the single-crystal Si and Sapphire substrates. The reflectivity curves were calculated using the IMD package \cite{alia04,win98}. The parameters used in the multilayer structural model (SLD values, interface roughness, magnetic dead layer thickness, etc) are adjusted for producing "realistic" spin-up and spin-down reflectivity profiles. This means that the main features of the simulated reflectivity profiles agree with the features usually measured with neutrons on SM produced with our coating process. As expected, the spin-up reflectivity curves are almost identical (in practice, high-frequency oscillations are completely washed out by the angular spread of the incident neutrons). A clear difference can be seen in the spin-down reflectivity (grey curves). No total reflection for spin-down neutrons is observed for the single-crystal Sapphire substrate. For the single-crystal Si substrate, spin-down neutrons are totally reflected at $m<0.4$ due to the positive step in SLD (see Table \ref{Table 1}). 

Using an "in-house" ray-tracing code in "Mathematica" and the reflectivity curves given above, we simulate the performance of "traditional" Si solid-state C bender, Si solid-state V-bender, and the proposed Sapphire V-bender. In all three cases, single-crystal wafers were assumed to be the substrates. In all simulations, we assumed a perfect plane parallel geometry of substrate plates, a plate thickness of $d=180 \mu m$ and a length of $L=50 mm$ for the C-bender and two plates $25 mm$ long each for the V-bender, see Figs. \ref{V},\ref{prototype}. In all calculations, we assumed a bending angle $20\%$ greater than that required for avoiding "direct view". This $20\%$ safety clearance takes into account the possible imperfection of the assembly and the shape of the substrate plates: $\gamma_c=10d/L$ for the C-bender and $\gamma_v=5d/L$ for the V-bender. With this bending angle, the cut-off wavelength of the C-bender is
\begin{eqnarray}
\lambda_{cut}= \frac{\sqrt{2d/R}}{m\Theta_{Ni}}=\sqrt{20}\frac{d/L}{m\Theta_{Ni}}\approx 0.29 nm,
\label{cut-off}
\end{eqnarray}
where $\theta_{Ni}=1.73\cdot 10^{-2}$ is the critical angle for neutrons with the wavelength of $1 nm$ and the natural Ni.

In our simulations, we also assume that the polarizer is installed downstream the exit of the H113 long ballistic neutron guide at ILL described in detail in \cite{abele06}. According to ref. \cite{abele06}, the angular distribution of cold neutrons at the exit of the H113 neutron guide is roughly the same as that expected at the exit of a straight neutron guide with walls coated with natural Ni; larger angles of reflection are suppressed by multiple reflections at reduced reflectivity. The gain by the value $m=2$ of the guide is thus mainly at short wavelengths. Another very important assumption is the full magnetic saturation of ferromagnetic layers in the SM coating, which means that the spin-flip reflectivity is set to zero, or $R_{+-}=R_{-+}=0$. Spin-flip scattering by the SM and the effect of the magnetic field strength on the polarizer performance will be considered in detail in the next chapter.

\begin{figure}
\includegraphics[width=1.05\linewidth]{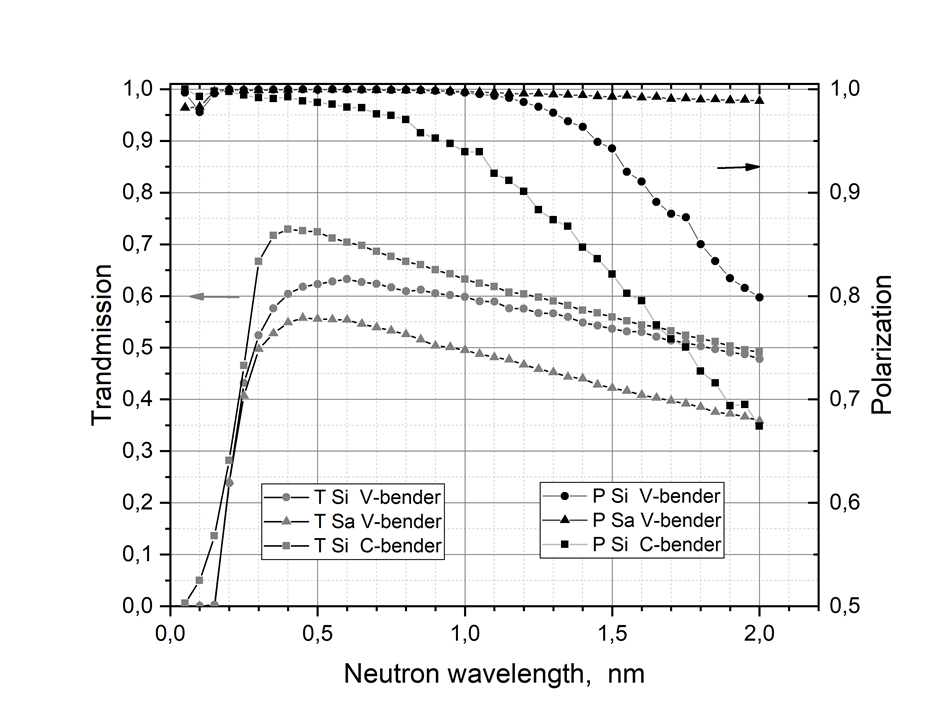}
\caption{\label{polar} Simulated polarization (black points) and transmission (grey points) for different solid-state Fe/SiN$_x$ polarizers.}
\end{figure}

\begin{figure}
\includegraphics[width=1.1\linewidth]{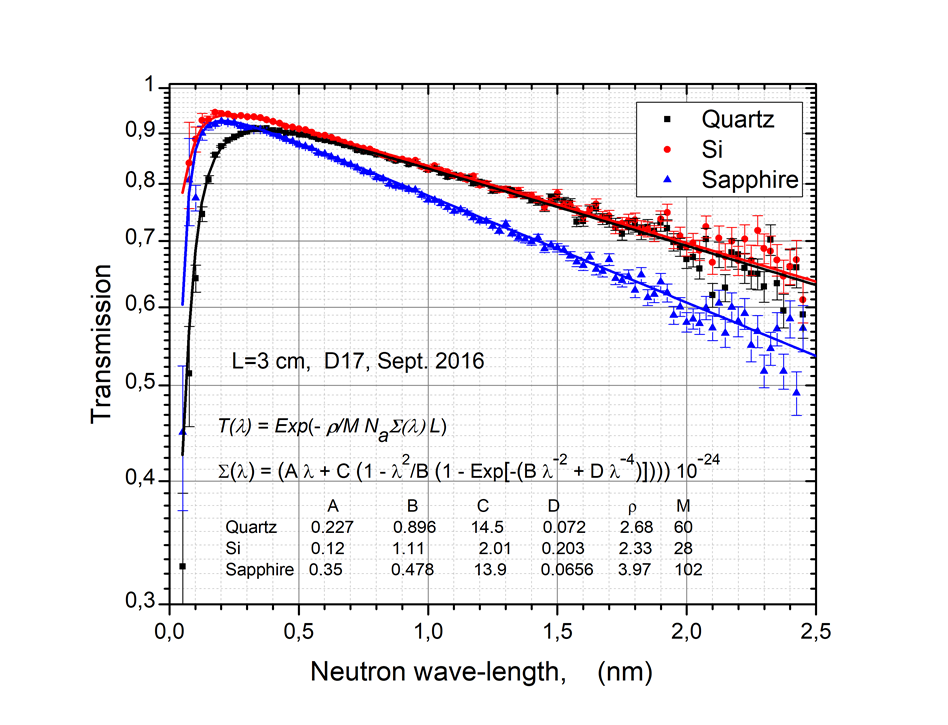}
\caption{\label{3 cm transmission} The neutron transmission of $3 cm$ thick single-crystal Quartz, Si, and Sapphire samples measured at D17 instrument at ILL.}
\end{figure}

Fig. \ref{polar} shows the simulation results for the Si solid-state C-bender (squares) as well as the Si solid-state V-bender (circles) and the Sapphire solid-state V-bender (triangles). Black points represent the polarizing power of the polarizer, and grey points stand for the polarizer transmission for spin-up neutrons. It can be seen that the choice of both the substrate material and geometry significantly affect the polarizer performance. A "classic" solid-state polarizer is the best choice if the transmission is the first priority while neutron polarization and wavelength band may not be the largest. The transition from C-bender to V-bender, while keeping Si substrates, significantly expands the useful wavelength band, but slightly reduces the transmission in the short-wavelength range. Finally, the proposed Sapphire V-bender dramatically improves polarization, showing only a $1\%$ decrease in polarization at long neutron wavelengths $\lambda=2 nm$. For long wavelengths, the difference in transmission is due to the difference in absorption, see Fig. \ref{3 cm transmission}, while for short wavelengths, the difference in transmission is dominated by the bender geometry effects. In practice, we expect that the issue of uncontrolled angular spread resulting from imperfect assembly of the mirror plates would be more serious for a C-bender than for a V-bender, whose geometry is simpler and allows in-situ controlling the inclination angle of each coated plate in the package. Considering this technical aspect, the difference in transmission at short wavelengths between the two geometries should be reduced with respect to Fig. \ref{polar}.

\section{\label{sec:level1}Depolarization of neutrons upon their reflection by SM}

As mentioned in the previous section, our Monte Carlo ray-tracing code as well as all other popular packages like VITESS \cite{VITESS} or McStas \cite{McStas} fully neglect the neutron spin-flip upon its reflection from SM. This is valid only if the local magnetization of the magnetic layers of the SM ideally follows the applied magnetic field for all domains. The general opinion is that a typical magnetic field applied to a SM of the order of several tens of $mT$ is sufficient to make the effects of spin-flip negligible. 

Nevertheless, whenever it has been carefully measured, all known SM polarizers have shown lower polarization than it was expected according to simulations, see, for example \cite{shah14,mish09,ino11}. The difference can be as high as a few percent even at the wavelength of the maximum expected polarization. This fact indicates that the assumption $R_{-+}=R_{+-}=0$ is overoptimistic. Spin-flip reflection is caused by the non-collinearity of magnetic moments in the SM coating and the non-collinearity of the applied magnetic field \textbf{H} and the mean field in the coating, \textbf{B}=\textbf{H}+4$\pi$\textbf{M}, where \textbf{M} is the mean magnetization in the region coherently illuminated by the neutron wave scattered at the coating ("local magnetization"). Assuming non-polarized incident neutron beam, the reflected beam polarization is \cite{plesh94}
\begin{eqnarray}
P[\chi]=\frac{R_{+}[0]-R_{-}[0]}{R_{+}[0]+R_{-}[0]}\cos{[\chi]}=P[0]\cos{[\chi]},
\label{depolarization}
\end{eqnarray}
where $\chi$ is the angle between the reflecting plane, in which the average coating magnetization is lying, and \textbf{H}.

In the matrix formalism, spin-dependent reflectivity $R_{\pm}[0]$ can be found from equation
\begin{eqnarray}
\left( 
\begin{array}{c}
R_{+}\\
R_{-}
\end{array} 
\right)
=
\left( 
\begin{array}{cc}
R_{++} & R_{-+}\\
R_{+-} & R_{--}
\end{array}
\right)
\frac{1}{2}
\left(
\begin{array}{c}
1\\
1
\end{array}
\right)
\
\label{matrix}.
\end{eqnarray}

Typically $R_{++}$ is the dominant term, ($R_{++}\gg R_{--},R_{-+}$); and $R_{-+}=R_{+-}$. Keeping only first order terms, polarization after a single reflection is given as follows
\begin{eqnarray}
P[0]\simeq 1-2\frac{R_{--}[0]+R_{+-}[0]}{R_{++}[0]}.
\label{first term}
\end{eqnarray}

Now consider two successive reflections from SM.
\begin{eqnarray}
\left( 
\begin{array}{c}
R_{+}\\
R_{-}
\end{array} 
\right)
=
{
\left(
\begin{array}{cc}
R_{++}^{j} & R_{-+}^{j}\\
R_{+-}^{j} & R_{--}^{j}
\end{array}
\right)
}
_{j=2}
{
\left(
\begin{array}{cc}
R_{++}^{j} & R_{-+}^{j}\\
R_{+-}^{j} & R_{--}^{j}
\end{array}
\right)
}
_{j=1}
\frac{1}{2}
\left(
\begin{array}{c}
1\\
1
\end{array}
\right)
\
\label{larger matrix},
\end{eqnarray}
where $j=1,2$ denotes first and second reflection. Again, using eqs. (\ref{depolarization},\ref{larger matrix}) and keeping only first order terms, the polarization after two successive reflections is
\begin{eqnarray}
P[0]\simeq1-2\frac{R_{+-}^{j}[0]}{R_{++}^{j}[0]}|_{j=2}.
\label{pol}
\end{eqnarray}

From eq. (\ref{pol}) it immediately follows that the polarization after the polarizer that provides multiple reflections is completely determined by the spin-flip at the last reflection. This situation is different from a single-reflection polarizer, where Eq. \ref{first term} applies and the final polarization is often dominated by the ratio $R_{--}/R_{++}$. This depolarization is neglected in all ray-tracing software. However, our previous studies \cite{klau13,klau16} show that this effect can be responsible for polarization losses of the percent level for applied fields of $10-100 mT$, which is often assumed to be large enough, while increasing the field further continuously decreases the losses caused by depolarization.

\section{\label{sec:level1}Concept validation}

To validate the concept of Sapphire V-bender, we performed tests with a prototype shown in Fig. \ref{prototype}. It consists of two stacks of single-crystal Sapphire plates $80x25 mm^2$ in size and $180 \mu$m thick, which were coated on both sides, at ILL, with $m=3.2$Fe/SiN$_{x}$ SM capped by a Si/Gd antireflecting and absorbing multilayer. Each stack consists of about 20 polarizing mirrors making the prototype polarizer cross-section of $80x3.6 mm^2$. Neutron polarization at the exit of the prototype was measured at PF1B instrument using a set of "opaque" cells filled in with polarized $^3$He \cite{klau13}. The $^3$He polarization was 75\%, the $^3$He pressure was $0.45, 0.8, 1.22, 1.7 bar$, the cell length along the neutron beam was $15 cm$. This set of "opaque" $^3$He cells provided more that $0.999$ analyzing power for the neutron wavelengths $>1.4, 0.8, 0.6, 0.4 nm$. The higher is $^3$He pressure the shorter is neutron wavelength with $0.999$ analyzing power. The time-of-flight technique was used to measure the neutron wavelength spectrum. The polarizer was installed after the neutron chopper without any angular collimation of the incident neutron beam.

\begin{figure}
\includegraphics[width=1.1\linewidth]{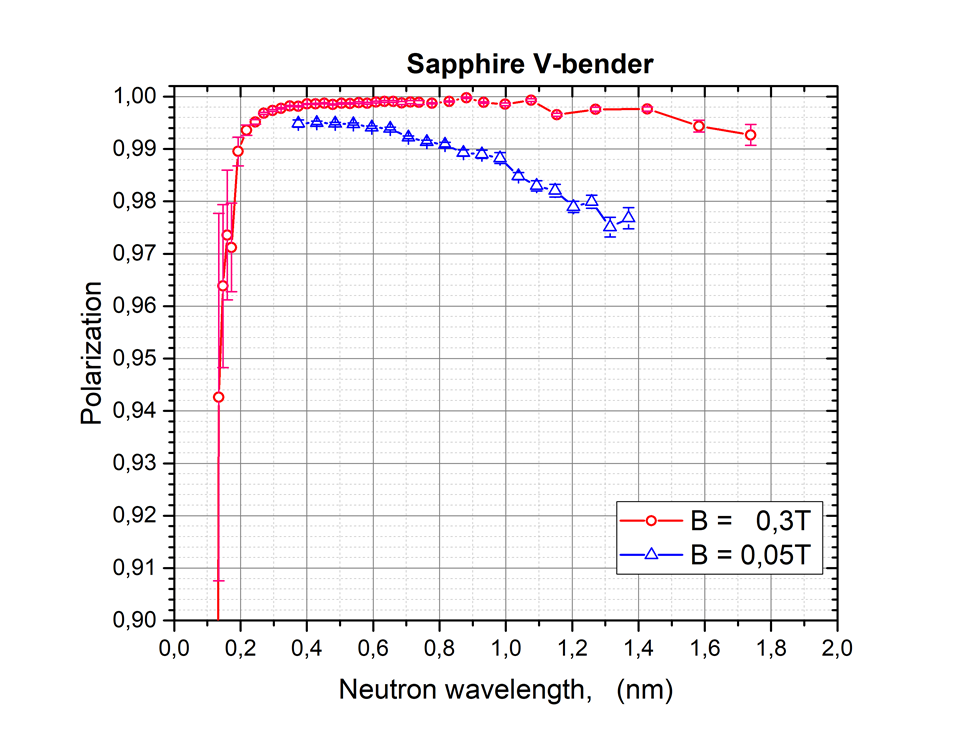}
\caption{\label{zig-zag} Measured neutron polarization downstream the Sapphire prototype V-bender measured with different strength of the applied magnetic field. Only neutrons having experienced double reflections are selected. No polarization correction was applied.}
\end{figure}

A specific feature of the V-bender is a quasi-discrete angular distribution in the outgoing neutron beam that results from different numbers of collisions of neutrons with SM in the polarizer. For a polarizer with larger cross-section, this effect is washed out due to the averaging over the polarizer width. For the prototype with a relatively small width (only $3.6 mm$), we were able to single out only neutrons experiencing two collisions in "zig-zag" geometry. Fig. \ref{zig-zag} shows the polarization data for neutrons with double-reflection trajectories.  

We performed two sets of measurements: with the strength of the applied magnetic field of only $50 mT$ (triangle points) and with a strong magnetic field of $300 mT$, see Fig. \ref{zig-zag}. The results obtained clearly demonstrate that the strength of the applied magnetic field, typical of most SM applications, is insufficient to suppress the spin-flip upon reflection. In order to suppress the depolarization, and accordingly expand the high-polarization bandwidth, it is necessary to apply a magnetic field about 10 times stronger than $50 mT$. 

To our knowledge, the polarization obtained with our prototype is the highest ever measured with cold neutrons of a broad wavelength band. The flipping ratio $N_{+}/N_{-}$ higher than $10^3$ was observed without any polarization correction. The polarization decrease for wavelengths below $0.3 nm$ is due to the decrease of $^3$He cell analyzing power at short wavelengths. 

\begin{figure}
\includegraphics[width=1.15\linewidth]{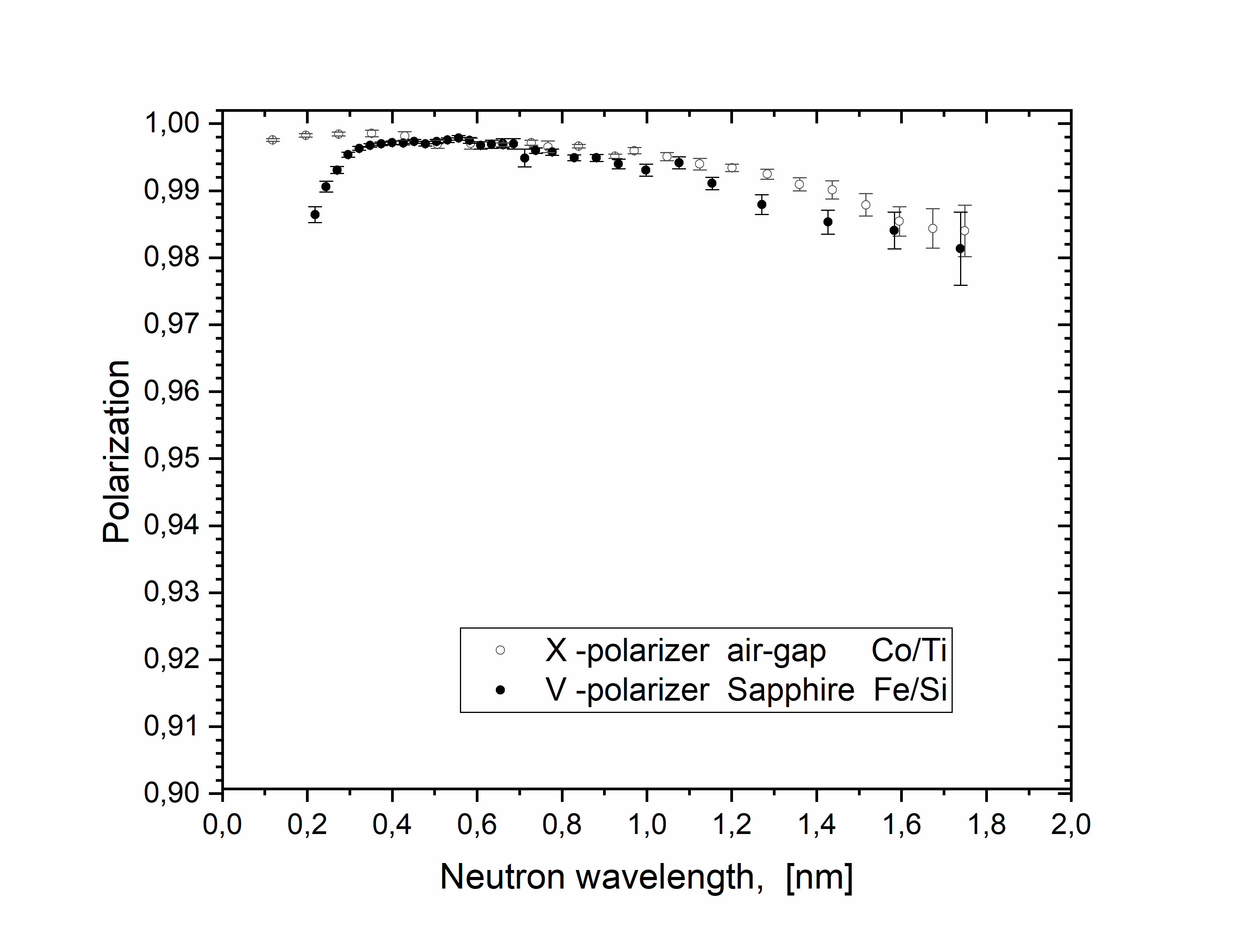}
\caption{\label{X} Neutron polarization as a function of wavelength for different polarizers. Open circles represent data for a $2 m$ long double Co/Ti polarizer in the X-SM geometry \cite{kreuz05,sol02}. Black circles show data for $6 cm$ long Fe/Si Sapphire V-bender without angular collimation, see text.}
\end{figure}

We also compare the polarization at the exit of a Sapphire V-bender Fe/Si prototype with the $6 cm$ total length for all transmitted neutron trajectories (whole outgoing beam) with the polarizing X-SM-bender described in detail in \cite{kreuz05} and used in \cite{schum07} to produce a neutron beam of ultra-high polarization, see Fig. \ref{X}. The Fe/Si Sapphire V-bender shows the polarization performance equivalent to that of the Co/Ti X-polarizer but 30 times more compact, free from the neutron induced degradation and activation of Co. For both polarizers, the beam mean polarization was measured to be $0.997$. For technical reasons, we were not able to measure directly the transmission probability but our simulations show a factor of 2 higher transmission for the Sapphire V-bender ($t\approx 50\%$, see Fig. \ref{polar}) than the transmission for the Co/Ti X-SM-bender ($t\approx 25\%$ \cite{kreuz05}) for spin-up neutrons. 

\section{\label{sec:level1}Magnetic system}

\begin{figure}
\includegraphics[width=\linewidth]{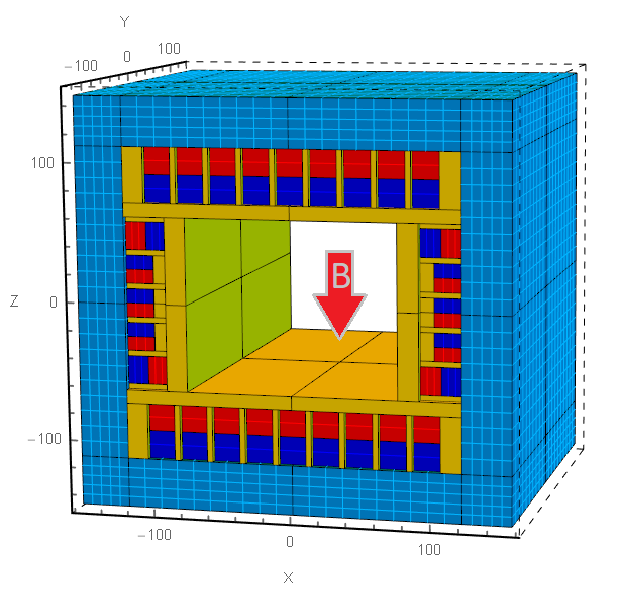}
\caption{\label{magnetic structure} A general sketch of the magnet for the SM polarizer of cold neutrons. Red color indicates the South pole and dense blue color shows the North pole. Dimensions are in $mm$.}
\end{figure}

As we have shown (Fig. \ref{zig-zag}), a sufficiently strong magnetic field ($B\geq 300 mT$) has to be applied to SMs located in the center of the magnetic system, in order to magnetize polarizing coatings and achieve a high polarization efficiency ($\approx 99.9\%$). Moreover, resulting from Eq. \ref{depolarization}, the angle $\chi$ has to be smaller than $2^{o}$ in the volume occupied by the polarizing stacks to suppress depolarization (the region of interest is $80x80x80 mm^{3}$). We designed a compact dipole magnet inspired by the original Halbach design \cite{hal80} that can be made from many identical permanent magnets. As a magnetic element, we chose relatively small rare-earth NbFeB magnets (size $20x20x10 mm^3$, grade N48H, $B_r=1.37 T$, $H_c=995 kA/m$). The small size of the magnetic element allows to avoid the occurrence of very large forces during the assembly process. Fig. \ref{magnetic structure} shows a sketch of the magnetic structure: the pure iron ARMCO yoke $40 mm$ thick is shown in light blue color, non-magnetic pieces defining positions of magnetic elements by yellow color, and NdFeB magnets by blue-red color. The total size of this structure is $\{L_x,L_y,L_z\}=\{312,270,294\} mm$, the magnet opening in the center is $\{L_x,L_y,L_z\}=\{152,270,124\} mm$, and the total weight is $130 kg$. The magnetic structure at the top and bottom of the magnet is composed of 9 bars each of 4 magnetic elements in the vertical direction (Z) and 12 elements in the longitudinal direction (Y). The magnetic structures at the left and right sides increase the strength and improve the field homogeneity in the region of interest.    

\begin{figure}
\includegraphics[width=\linewidth]{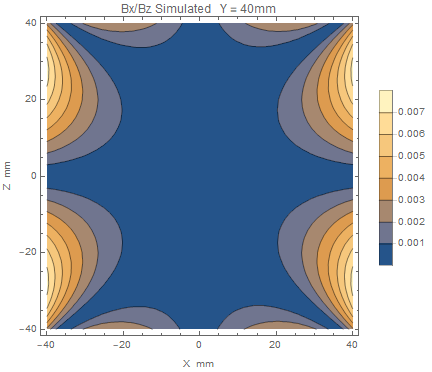}
\caption{\label{Z field} A calculated map of the relative transverse component $B_{x}/B_{z}$ in the plane shifted by $40 mm$ along the neutron beam axis from the center of the magnet (the worse plane where the Si plates will sit).}
\end{figure}

\begin{figure}
\includegraphics[width=0.9\linewidth]{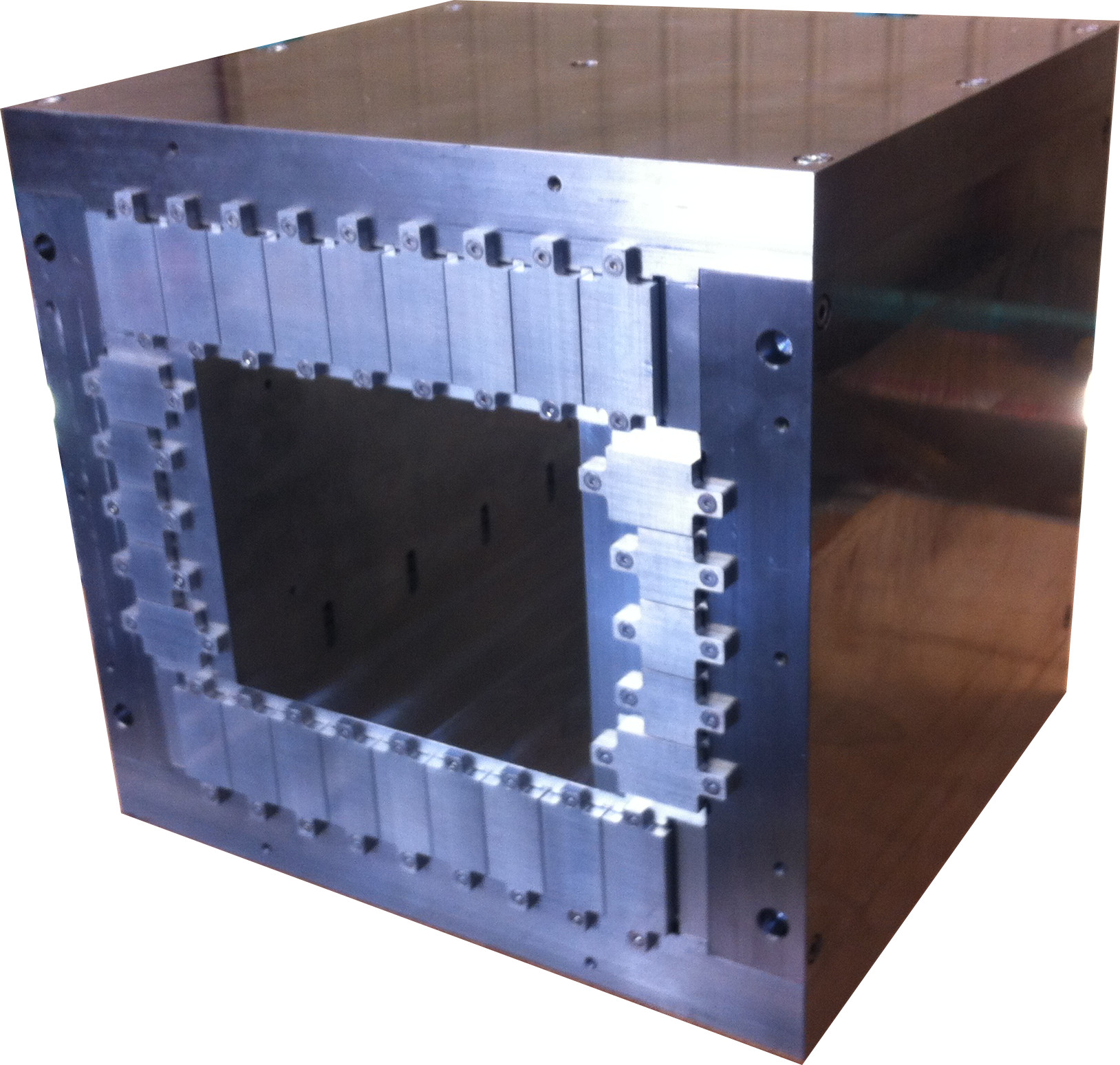}
\caption{\label{magnetic photo} A photo of the manufactured magnet for the new PF1B polarizer.}
\end{figure}

\begin{figure}
\includegraphics[width=\linewidth]{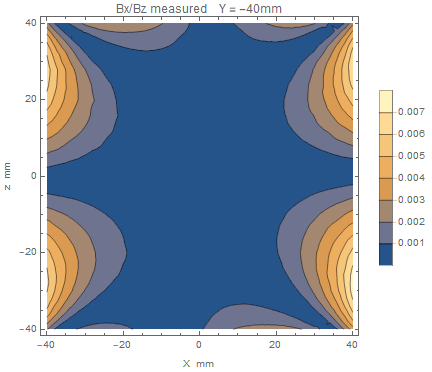}
\caption{\label{Y field} A measured map of the relative transverse component of the field in the aperture of the magnet shown in Fig. \ref{magnetic photo} (compare to the calculated map shown in Fig. \ref{Z field}).}
\end{figure}

To simulate the magnetic field produced by the large number of small magnets, we used the "Mathematica" package "Radia" dedicated to 3D magneto-static computations \cite{radia,elle97}. We found that the horizontal distance along X between side structures is a critical parameter that defines the field homogeneity within the aperture. Fig. \ref{Z field} shows a map of the calculated relative transverse component $\chi\approx B_{x}/B_{z}$ in the magnet aperture for the optimized magnet structure. As can be seen, the requirement that $\chi<2^{o}$ is widely met. The strength of the central magnetic field is $B_{z0}=0.374 T$. Fig. \ref{magnetic photo} shows a photo of the manufactured magnetic system, and Fig. \ref{Y field} presents the measured $B_{x}/B_{z}$ map.

We found an excellent agreement between the simulated and measured fields. A very useful feature of the magnet is very small stray fields. We measured only a few Gauss at the distance of $10 cm$ away from the yoke with $B\approx 0.37 T$ in the aperture of the magnet.

\section{\label{sec:level1}Conclusion}

Solid-state neutron-optics devices (polarizers, collimators, deflectors...) have many advantages over more traditional air-gap devices. The most important are 5-10 times smaller size along the neutron beam direction, the possibility to apply stronger magnetic field to magnetize SMs, and the possibility to use FeSi SM coatings that can well withstand very high neutron fluences. An important condition for a good solid-state polarizer is the matching of the substrate SLD with the SM coating SLD ($\rho_{Fe}^{-}$ for FeSi SM on Si) for the spin-down component of neutron polarization. For traditional FeSi SM on Si substrate, this step is positive, which leads to a significant degradation of the polarizer performance in the small Q region. To solve this problem, we proposed in our previous publication \cite{petukh16} to replace single-crystal Si substrates by single-crystal Sapphire substrates. The latter show a negative SLD step for the neutron spin-down component at the interface with Fe and, therefore, avoid the total reflection regime in the small Q region. Careful analysis of the factors limiting the polarizer performance allowed us to formulate the concept of Sapphire V-bender. Our ray-tracing simulations show a dramatic improvement in the polarizing power ($P=0.999$ for $\lambda =0.3-1.2 nm$ and only $1\%$ less at the wavelength of $\lambda = 2.0 nm$). The price for improving polarization is $15\%$ lower absolute transmission, see Fig. \ref{polar}, due to the higher absorption in Sapphire of the neutrons with longer wavelengths, as well as due the V-bender geometry for neutrons with shorter wavelengths. Imperfections in the plane-parallel geometry of the substrates, dust particles falling between the substrates, stress-induced deformations of the plates, all this leads to an additional angular spread of the reflecting plates, and therefore leads to less than expected efficiency of polarizers, which was often observed \cite{shah14,mish09,ino11}. The simple geometry of each stack of the V-Bender simplifies the optical in-situ control of the tilting angle of each coated substrate during the assembly procedure. We also noticed that all known ray-tracing packages assume no neutron depolarization upon their reflection from SMs, $R_{-+}=R_{+-}=0$. As a result, the simulated polarization is always higher than the measured one. For a magnetizing field of below $100 mT$, the depolarization can be as high as a few percent depending on the $m$-value of the SM coating \cite{klau16}. A significantly higher magnetizing field is needed to suppress the depolarization. Simultaneously, special care has to be devoted to keeping the applied field direction strictly parallel to the mirror surfaces in order to avoid depolarization. To validate the proposed new concept of Sapphire V-bender, we built a prototype device. The Sapphire V-bender prototype was tested at the PF1B instrument at ILL using "opaque" optically polarized $^{3}$He (with the analyzing power of higher than 0.999). We compared the polarizer performance for the applied magnetic fields of $50 mT$ and $300 mT$. Only for the high field ($300 mT$), we found an excellent agreement between the simulated and measured polarization. To our knowledge, this is the first time when the simulated and measured polarization coincided in a broad wavelength band ($0.3-2.0 nm$) demonstrating ultra-high polarization $P>0.999$ for $\lambda\approx 0.3-1.2 nm$ and only $1\%$ less at the wavelength $\lambda=2 nm$. Without any additional collimation, the Sapphire V-bender demonstrates 2 times higher transmission and the polarization performance equivalent to that of a double polarizer in the X-SM geometry. The polarization averaged over the outgoing neutron capture spectrum is $0.997$. A full-scale polarizer for the PF1B instrument at ILL, based on the new concept of solid-state Sapphire V-bender, is currently being produced. We plan to complete this project by the end of 2019.   

\begin{acknowledgments}
We express our gratitude to our technicians Didier Berruer and Pascal Mouveau who produced many technical components of the experiment and to Guillaume Delphin, Guillaume Delphin, Amandine Vittoz, Florian Philit and Vincent Gaignon for the production of super-mirrors by magnetron sputtering. We thank Thomas Saerbeck for useful transmission and reflectivity measurements on substrate materials on D17 instrument.
\end{acknowledgments}

\bibliography{aipsamp}

\providecommand{\noopsort}[1]{}\providecommand{\singleletter}[1]{#1}%
\begin{thebibliography}{41}%
\makeatletter
\providecommand \@ifxundefined [1]{%
 \@ifx{#1\undefined}
}%
\providecommand \@ifnum [1]{%
 \ifnum #1\expandafter \@firstoftwo
 \else \expandafter \@secondoftwo
 \fi
}%
\providecommand \@ifx [1]{%
 \ifx #1\expandafter \@firstoftwo
 \else \expandafter \@secondoftwo
 \fi
}%
\providecommand \natexlab [1]{#1}%
\providecommand \enquote  [1]{``#1''}%
\providecommand \bibnamefont  [1]{#1}%
\providecommand \bibfnamefont [1]{#1}%
\providecommand \citenamefont [1]{#1}%
\providecommand \href@noop [0]{\@secondoftwo}%
\providecommand \href [0]{\begingroup \@sanitize@url \@href}%
\providecommand \@href[1]{\@@startlink{#1}\@@href}%
\providecommand \@@href[1]{\endgroup#1\@@endlink}%
\providecommand \@sanitize@url [0]{\catcode `\\12\catcode `\$12\catcode
  `\&12\catcode `\#12\catcode `\^12\catcode `\_12\catcode `\%12\relax}%
\providecommand \@@startlink[1]{}%
\providecommand \@@endlink[0]{}%
\providecommand \url  [0]{\begingroup\@sanitize@url \@url }%
\providecommand \@url [1]{\endgroup\@href {#1}{\urlprefix }}%
\providecommand \urlprefix  [0]{URL }%
\providecommand \Eprint [0]{\href }%
\providecommand \doibase [0]{http://dx.doi.org/}%
\providecommand \selectlanguage [0]{\@gobble}%
\providecommand \bibinfo  [0]{\@secondoftwo}%
\providecommand \bibfield  [0]{\@secondoftwo}%
\providecommand \translation [1]{[#1]}%
\providecommand \BibitemOpen [0]{}%
\providecommand \bibitemStop [0]{}%
\providecommand \bibitemNoStop [0]{.\EOS\space}%
\providecommand \EOS [0]{\spacefactor3000\relax}%
\providecommand \BibitemShut  [1]{\csname bibitem#1\endcsname}%
\let\auto@bib@innerbib\@empty
\bibitem [{\citenamefont {Petukhov}\ \emph {et~al.}(2016)\citenamefont
  {Petukhov}, \citenamefont {Nesvizhevsky}, \citenamefont {Bigault},
  \citenamefont {Courtois}, \citenamefont {Jullien},\ and\ \citenamefont
  {Soldner}}]{petukh16}%
  \BibitemOpen
  \bibfield  {author} {\bibinfo {author} {\bibfnamefont {A.~K.}\ \bibnamefont
  {Petukhov}}, \bibinfo {author} {\bibfnamefont {V.~V.}\ \bibnamefont
  {Nesvizhevsky}}, \bibinfo {author} {\bibfnamefont {T.}~\bibnamefont
  {Bigault}}, \bibinfo {author} {\bibfnamefont {P.}~\bibnamefont {Courtois}},
  \bibinfo {author} {\bibfnamefont {D.}~\bibnamefont {Jullien}}, \ and\
  \bibinfo {author} {\bibfnamefont {T.}~\bibnamefont {Soldner}},\ }\href@noop
  {} {\bibfield  {journal} {\bibinfo  {journal} {Nucl. Instr. Meth. A}\
  }\textbf {\bibinfo {volume} {838}},\ \bibinfo {pages} {33} (\bibinfo {year}
  {2016})}\BibitemShut {NoStop}%
\bibitem [{\citenamefont {Kreuz}\ \emph {et~al.}(2005)\citenamefont {Kreuz},
  \citenamefont {Nesvizhevsky}, \citenamefont {Petukhov},\ and\ \citenamefont
  {Soldner}}]{kreuz05}%
  \BibitemOpen
  \bibfield  {author} {\bibinfo {author} {\bibfnamefont {M.}~\bibnamefont
  {Kreuz}}, \bibinfo {author} {\bibfnamefont {V.~V.}\ \bibnamefont
  {Nesvizhevsky}}, \bibinfo {author} {\bibfnamefont {A.}~\bibnamefont
  {Petukhov}}, \ and\ \bibinfo {author} {\bibfnamefont {T.}~\bibnamefont
  {Soldner}},\ }\href@noop {} {\bibfield  {journal} {\bibinfo  {journal} {Nucl.
  Instr. Meth. A}\ }\textbf {\bibinfo {volume} {547}},\ \bibinfo {pages} {583}
  (\bibinfo {year} {2005})}\BibitemShut {NoStop}%
\bibitem [{\citenamefont {Abele}\ \emph {et~al.}(2006)\citenamefont {Abele},
  \citenamefont {Dubbers}, \citenamefont {Hase}, \citenamefont {Klein},
  \citenamefont {Knopfler}, \citenamefont {Kreuz}, \citenamefont {Lauer},
  \citenamefont {Markisch}, \citenamefont {Mund}, \citenamefont {Nesvizhevsky},
  \citenamefont {Petukhov}, \citenamefont {Schmidt}, \citenamefont {Shumann},\
  and\ \citenamefont {Soldner}}]{abele06}%
  \BibitemOpen
  \bibfield  {author} {\bibinfo {author} {\bibfnamefont {H.}~\bibnamefont
  {Abele}}, \bibinfo {author} {\bibfnamefont {D.}~\bibnamefont {Dubbers}},
  \bibinfo {author} {\bibfnamefont {H.}~\bibnamefont {Hase}}, \bibinfo {author}
  {\bibfnamefont {M.}~\bibnamefont {Klein}}, \bibinfo {author} {\bibfnamefont
  {A.}~\bibnamefont {Knopfler}}, \bibinfo {author} {\bibfnamefont
  {M.}~\bibnamefont {Kreuz}}, \bibinfo {author} {\bibfnamefont
  {T.}~\bibnamefont {Lauer}}, \bibinfo {author} {\bibfnamefont
  {B.}~\bibnamefont {Markisch}}, \bibinfo {author} {\bibfnamefont
  {D.}~\bibnamefont {Mund}}, \bibinfo {author} {\bibfnamefont {V.~V.}\
  \bibnamefont {Nesvizhevsky}}, \bibinfo {author} {\bibfnamefont
  {A.}~\bibnamefont {Petukhov}}, \bibinfo {author} {\bibfnamefont
  {C.}~\bibnamefont {Schmidt}}, \bibinfo {author} {\bibfnamefont
  {M.}~\bibnamefont {Shumann}}, \ and\ \bibinfo {author} {\bibfnamefont
  {T.}~\bibnamefont {Soldner}},\ }\href@noop {} {\bibfield  {journal} {\bibinfo
   {journal} {Nucl. Instr. Meth. A}\ }\textbf {\bibinfo {volume} {562}},\
  \bibinfo {pages} {40} (\bibinfo {year} {2006})}\BibitemShut {NoStop}%
\bibitem [{\citenamefont {Vesna}\ \emph {et~al.}(2017)\citenamefont {Vesna},
  \citenamefont {Gledenov}, \citenamefont {Nesvizhevsky}, \citenamefont
  {Petoukhov}, \citenamefont {Sedyshev}, \citenamefont {Soldner}, \citenamefont
  {Zimmer},\ and\ \citenamefont {Shulgina}}]{vesna08}%
  \BibitemOpen
  \bibfield  {author} {\bibinfo {author} {\bibfnamefont {V.~A.}\ \bibnamefont
  {Vesna}}, \bibinfo {author} {\bibfnamefont {Y.~M.}\ \bibnamefont {Gledenov}},
  \bibinfo {author} {\bibfnamefont {V.~V.}\ \bibnamefont {Nesvizhevsky}},
  \bibinfo {author} {\bibfnamefont {A.~K.}\ \bibnamefont {Petoukhov}}, \bibinfo
  {author} {\bibfnamefont {P.~V.}\ \bibnamefont {Sedyshev}}, \bibinfo {author}
  {\bibfnamefont {T.}~\bibnamefont {Soldner}}, \bibinfo {author} {\bibfnamefont
  {O.}~\bibnamefont {Zimmer}}, \ and\ \bibinfo {author} {\bibfnamefont {E.~V.}\
  \bibnamefont {Shulgina}},\ }\href@noop {} {\bibfield  {journal} {\bibinfo
  {journal} {Phys. Lett. B}\ }\textbf {\bibinfo {volume} {769}},\ \bibinfo
  {pages} {111} (\bibinfo {year} {2017})}\BibitemShut {NoStop}%
\bibitem [{\citenamefont {Gledenov}\ \emph {et~al.}(2017)\citenamefont
  {Gledenov}, \citenamefont {Nesvizhevsky}, \citenamefont {Sedyshev},
  \citenamefont {Shul'gina}, \citenamefont {Czalanski},\ and\ \citenamefont
  {Vesna}}]{vesna17}%
  \BibitemOpen
  \bibfield  {author} {\bibinfo {author} {\bibfnamefont {Y.~M.}\ \bibnamefont
  {Gledenov}}, \bibinfo {author} {\bibfnamefont {V.~V.}\ \bibnamefont
  {Nesvizhevsky}}, \bibinfo {author} {\bibfnamefont {P.~V.}\ \bibnamefont
  {Sedyshev}}, \bibinfo {author} {\bibfnamefont {E.~V.}\ \bibnamefont
  {Shul'gina}}, \bibinfo {author} {\bibfnamefont {P.}~\bibnamefont
  {Czalanski}}, \ and\ \bibinfo {author} {\bibfnamefont {V.~A.}\ \bibnamefont
  {Vesna}},\ }\href@noop {} {\bibfield  {journal} {\bibinfo  {journal} {Phys.
  Rev. C}\ }\textbf {\bibinfo {volume} {77}},\ \bibinfo {pages} {035501}
  (\bibinfo {year} {2017})}\BibitemShut {NoStop}%
\bibitem [{\citenamefont {Goennenwein}\ \emph {et~al.}(2007)\citenamefont
  {Goennenwein}, \citenamefont {Mutterer}, \citenamefont {Gagarski},
  \citenamefont {Guseva}, \citenamefont {Petrov}, \citenamefont {Sokolov},
  \citenamefont {Zavarukhina}, \citenamefont {Gusev}, \citenamefont {von
  Kalben}, \citenamefont {Nesvizhevski},\ and\ \citenamefont
  {Soldner}}]{goen07}%
  \BibitemOpen
  \bibfield  {author} {\bibinfo {author} {\bibfnamefont {F.}~\bibnamefont
  {Goennenwein}}, \bibinfo {author} {\bibfnamefont {M.}~\bibnamefont
  {Mutterer}}, \bibinfo {author} {\bibfnamefont {A.}~\bibnamefont {Gagarski}},
  \bibinfo {author} {\bibfnamefont {I.}~\bibnamefont {Guseva}}, \bibinfo
  {author} {\bibfnamefont {G.}~\bibnamefont {Petrov}}, \bibinfo {author}
  {\bibfnamefont {V.}~\bibnamefont {Sokolov}}, \bibinfo {author} {\bibfnamefont
  {T.}~\bibnamefont {Zavarukhina}}, \bibinfo {author} {\bibfnamefont
  {Y.}~\bibnamefont {Gusev}}, \bibinfo {author} {\bibfnamefont
  {J.}~\bibnamefont {von Kalben}}, \bibinfo {author} {\bibfnamefont
  {V.}~\bibnamefont {Nesvizhevski}}, \ and\ \bibinfo {author} {\bibfnamefont
  {T.}~\bibnamefont {Soldner}},\ }\href@noop {} {\bibfield  {journal} {\bibinfo
   {journal} {Phys. Lett. B}\ }\textbf {\bibinfo {volume} {652}},\ \bibinfo
  {pages} {13} (\bibinfo {year} {2007})}\BibitemShut {NoStop}%
\bibitem [{\citenamefont {Gagarski}\ \emph {et~al.}(2016)\citenamefont
  {Gagarski}, \citenamefont {Gonnenwein}, \citenamefont {Guseva}, \citenamefont
  {Jesinger}, \citenamefont {Kopatch}, \citenamefont {Kuzmina}, \citenamefont
  {Lelievre-Berna}, \citenamefont {Mutterer}, \citenamefont {Nesvizhevsky},
  \citenamefont {Petrov}, \citenamefont {Soldner}, \citenamefont {Tiourine},
  \citenamefont {Trzaska},\ and\ \citenamefont {Zavarukhina}}]{gagar16}%
  \BibitemOpen
  \bibfield  {author} {\bibinfo {author} {\bibfnamefont {A.}~\bibnamefont
  {Gagarski}}, \bibinfo {author} {\bibfnamefont {F.}~\bibnamefont
  {Gonnenwein}}, \bibinfo {author} {\bibfnamefont {I.}~\bibnamefont {Guseva}},
  \bibinfo {author} {\bibfnamefont {P.}~\bibnamefont {Jesinger}}, \bibinfo
  {author} {\bibfnamefont {Y.}~\bibnamefont {Kopatch}}, \bibinfo {author}
  {\bibfnamefont {T.}~\bibnamefont {Kuzmina}}, \bibinfo {author} {\bibfnamefont
  {E.}~\bibnamefont {Lelievre-Berna}}, \bibinfo {author} {\bibfnamefont
  {M.}~\bibnamefont {Mutterer}}, \bibinfo {author} {\bibfnamefont
  {V.}~\bibnamefont {Nesvizhevsky}}, \bibinfo {author} {\bibfnamefont
  {G.}~\bibnamefont {Petrov}}, \bibinfo {author} {\bibfnamefont
  {T.}~\bibnamefont {Soldner}}, \bibinfo {author} {\bibfnamefont
  {G.}~\bibnamefont {Tiourine}}, \bibinfo {author} {\bibfnamefont {W.~H.}\
  \bibnamefont {Trzaska}}, \ and\ \bibinfo {author} {\bibfnamefont
  {T.}~\bibnamefont {Zavarukhina}},\ }\href@noop {} {\bibfield  {journal}
  {\bibinfo  {journal} {Phys. Rev. C}\ }\textbf {\bibinfo {volume} {93}},\
  \bibinfo {pages} {054619} (\bibinfo {year} {2016})}\BibitemShut {NoStop}%
\bibitem [{\citenamefont {Mund}\ \emph {et~al.}(2013)\citenamefont {Mund},
  \citenamefont {Markisch}, \citenamefont {Diesseroth}, \citenamefont
  {Krempel}, \citenamefont {Schumann}, \citenamefont {Abele}, \citenamefont
  {Petukhov},\ and\ \citenamefont {Soldner}}]{schum07}%
  \BibitemOpen
  \bibfield  {author} {\bibinfo {author} {\bibfnamefont {D.}~\bibnamefont
  {Mund}}, \bibinfo {author} {\bibfnamefont {B.}~\bibnamefont {Markisch}},
  \bibinfo {author} {\bibfnamefont {M.}~\bibnamefont {Diesseroth}}, \bibinfo
  {author} {\bibfnamefont {J.}~\bibnamefont {Krempel}}, \bibinfo {author}
  {\bibfnamefont {M.}~\bibnamefont {Schumann}}, \bibinfo {author}
  {\bibfnamefont {H.}~\bibnamefont {Abele}}, \bibinfo {author} {\bibfnamefont
  {A.}~\bibnamefont {Petukhov}}, \ and\ \bibinfo {author} {\bibfnamefont
  {T.}~\bibnamefont {Soldner}},\ }\href@noop {} {\bibfield  {journal} {\bibinfo
   {journal} {Phys. Rev. Lett.}\ }\textbf {\bibinfo {volume} {110}},\ \bibinfo
  {pages} {172502} (\bibinfo {year} {2013})}\BibitemShut {NoStop}%
\bibitem [{\citenamefont {Mezei}\ and\ \citenamefont
  {Dagleish}(1977)}]{mezei77}%
  \BibitemOpen
  \bibfield  {author} {\bibinfo {author} {\bibfnamefont {F.}~\bibnamefont
  {Mezei}}\ and\ \bibinfo {author} {\bibfnamefont {P.~A.}\ \bibnamefont
  {Dagleish}},\ }\href@noop {} {\bibfield  {journal} {\bibinfo  {journal}
  {Commun. Phys.}\ }\textbf {\bibinfo {volume} {2}},\ \bibinfo {pages} {41}
  (\bibinfo {year} {1977})}\BibitemShut {NoStop}%
\bibitem [{\citenamefont {Drabkin}\ \emph {et~al.}(1977)\citenamefont
  {Drabkin}, \citenamefont {Okorokov}, \citenamefont {Shchebetov},
  \citenamefont {Borovikova}, \citenamefont {Gukasov}, \citenamefont {Korneev},
  \citenamefont {Kudryashov},\ and\ \citenamefont {Runov}}]{drab77}%
  \BibitemOpen
  \bibfield  {author} {\bibinfo {author} {\bibfnamefont {G.~M.}\ \bibnamefont
  {Drabkin}}, \bibinfo {author} {\bibfnamefont {A.~I.}\ \bibnamefont
  {Okorokov}}, \bibinfo {author} {\bibfnamefont {A.~F.}\ \bibnamefont
  {Shchebetov}}, \bibinfo {author} {\bibfnamefont {N.~V.}\ \bibnamefont
  {Borovikova}}, \bibinfo {author} {\bibfnamefont {A.~G.}\ \bibnamefont
  {Gukasov}}, \bibinfo {author} {\bibfnamefont {D.~A.}\ \bibnamefont
  {Korneev}}, \bibinfo {author} {\bibfnamefont {V.~A.}\ \bibnamefont
  {Kudryashov}}, \ and\ \bibinfo {author} {\bibfnamefont {V.~V.}\ \bibnamefont
  {Runov}},\ }\href@noop {} {\bibfield  {journal} {\bibinfo  {journal} {J.
  Techn. Phys.}\ }\textbf {\bibinfo {volume} {47}},\ \bibinfo {pages} {203}
  (\bibinfo {year} {1977})}\BibitemShut {NoStop}%
\bibitem [{\citenamefont {Schaerpf}(1986)}]{schaer86}%
  \BibitemOpen
  \bibfield  {author} {\bibinfo {author} {\bibfnamefont {O.}~\bibnamefont
  {Schaerpf}},\ }\href@noop {} {\bibfield  {journal} {\bibinfo  {journal}
  {Phys. B}\ }\textbf {\bibinfo {volume} {156}},\ \bibinfo {pages} {639}
  (\bibinfo {year} {1986})}\BibitemShut {NoStop}%
\bibitem [{\citenamefont {Majkrzak}(1995)}]{majk95}%
  \BibitemOpen
  \bibfield  {author} {\bibinfo {author} {\bibfnamefont {C.~F.}\ \bibnamefont
  {Majkrzak}},\ }\href@noop {} {\bibfield  {journal} {\bibinfo  {journal}
  {Phys. B}\ }\textbf {\bibinfo {volume} {213}},\ \bibinfo {pages} {904}
  (\bibinfo {year} {1995})}\BibitemShut {NoStop}%
\bibitem [{\citenamefont {Maruyama}\ \emph {et~al.}(2007)\citenamefont
  {Maruyama}, \citenamefont {Yamazaki}, \citenamefont {Ebisawa}, \citenamefont
  {Hino},\ and\ \citenamefont {Soyama}}]{maru07}%
  \BibitemOpen
  \bibfield  {author} {\bibinfo {author} {\bibfnamefont {R.}~\bibnamefont
  {Maruyama}}, \bibinfo {author} {\bibfnamefont {T.}~\bibnamefont {Yamazaki}},
  \bibinfo {author} {\bibfnamefont {T.}~\bibnamefont {Ebisawa}}, \bibinfo
  {author} {\bibfnamefont {M.}~\bibnamefont {Hino}}, \ and\ \bibinfo {author}
  {\bibfnamefont {K.}~\bibnamefont {Soyama}},\ }\href@noop {} {\bibfield
  {journal} {\bibinfo  {journal} {Sol. Thin Film.}\ }\textbf {\bibinfo {volume}
  {515}},\ \bibinfo {pages} {5704} (\bibinfo {year} {2007})}\BibitemShut
  {NoStop}%
\bibitem [{\citenamefont {Krist}\ \emph {et~al.}(2008)\citenamefont {Krist},
  \citenamefont {Teichert}, \citenamefont {Mezei},\ and\ \citenamefont
  {Rosta}}]{krist08}%
  \BibitemOpen
  \bibfield  {author} {\bibinfo {author} {\bibfnamefont {T.}~\bibnamefont
  {Krist}}, \bibinfo {author} {\bibfnamefont {A.}~\bibnamefont {Teichert}},
  \bibinfo {author} {\bibfnamefont {F.}~\bibnamefont {Mezei}}, \ and\ \bibinfo
  {author} {\bibfnamefont {L.}~\bibnamefont {Rosta}},\ }\href@noop {} {\emph
  {\bibinfo {title} {Modern developments in X-ray and neutron optics, pp.
  355-370, ISBN 978-3-540-74560-0}}}\ (\bibinfo  {publisher} {Springer-Verlag,
  Berlin, Heidelberg},\ \bibinfo {year} {2008})\BibitemShut {NoStop}%
\bibitem [{\citenamefont {Mezei}(1989)}]{mezei89}%
  \BibitemOpen
  \bibfield  {author} {\bibinfo {author} {\bibfnamefont {F.}~\bibnamefont
  {Mezei}},\ }\href@noop {} {\emph {\bibinfo {title} {SPIE Proc., vol. 983, p.
  10, C. F. Majkrzak (Ed.), The Film Neutron Optical Device}}}\ (\bibinfo
  {publisher} {Bermingham, WA},\ \bibinfo {year} {1989})\BibitemShut {NoStop}%
\bibitem [{\citenamefont {Elsenhans}\ \emph {et~al.}(1994)\citenamefont
  {Elsenhans}, \citenamefont {Boni}, \citenamefont {Friedli}, \citenamefont
  {Grimmer}, \citenamefont {Buffat}, \citenamefont {Leifer}, \citenamefont
  {Sochtig},\ and\ \citenamefont {Anderson}}]{and94}%
  \BibitemOpen
  \bibfield  {author} {\bibinfo {author} {\bibfnamefont {O.}~\bibnamefont
  {Elsenhans}}, \bibinfo {author} {\bibfnamefont {P.}~\bibnamefont {Boni}},
  \bibinfo {author} {\bibfnamefont {H.~P.}\ \bibnamefont {Friedli}}, \bibinfo
  {author} {\bibfnamefont {H.}~\bibnamefont {Grimmer}}, \bibinfo {author}
  {\bibfnamefont {P.}~\bibnamefont {Buffat}}, \bibinfo {author} {\bibfnamefont
  {K.}~\bibnamefont {Leifer}}, \bibinfo {author} {\bibfnamefont
  {J.}~\bibnamefont {Sochtig}}, \ and\ \bibinfo {author} {\bibfnamefont
  {I.~S.}\ \bibnamefont {Anderson}},\ }\href@noop {} {\bibfield  {journal}
  {\bibinfo  {journal} {Thin Solid Films}\ }\textbf {\bibinfo {volume} {246}},\
  \bibinfo {pages} {110} (\bibinfo {year} {1994})}\BibitemShut {NoStop}%
\bibitem [{\citenamefont {Courtois}\ \emph {et~al.}(2013)\citenamefont
  {Courtois}, \citenamefont {Bigault}, \citenamefont {Gaignon}, \citenamefont
  {Vottoz}, \citenamefont {Delphin},\ and\ \citenamefont {Bourgault}}]{cour13}%
  \BibitemOpen
  \bibfield  {author} {\bibinfo {author} {\bibfnamefont {P.}~\bibnamefont
  {Courtois}}, \bibinfo {author} {\bibfnamefont {T.}~\bibnamefont {Bigault}},
  \bibinfo {author} {\bibfnamefont {V.}~\bibnamefont {Gaignon}}, \bibinfo
  {author} {\bibfnamefont {A.}~\bibnamefont {Vottoz}}, \bibinfo {author}
  {\bibfnamefont {G.}~\bibnamefont {Delphin}}, \ and\ \bibinfo {author}
  {\bibfnamefont {D.}~\bibnamefont {Bourgault}},\ }\href@noop {} {\bibfield
  {journal} {\bibinfo  {journal} {ILL Annual Report}\ ,\ \bibinfo {pages} {82}}
  (\bibinfo {year} {2013})}\BibitemShut {NoStop}%
\bibitem [{\citenamefont {Soldner}, \citenamefont {Petukhov},\ and\
  \citenamefont {Plonka}(2002)}]{sol02}%
  \BibitemOpen
  \bibfield  {author} {\bibinfo {author} {\bibfnamefont {T.}~\bibnamefont
  {Soldner}}, \bibinfo {author} {\bibfnamefont {A.}~\bibnamefont {Petukhov}}, \
  and\ \bibinfo {author} {\bibfnamefont {C.}~\bibnamefont {Plonka}},\
  }\href@noop {} {}\bibinfo {type} {Technical Report}\ \bibinfo {number}
  {ILL03S010T}\ (\bibinfo  {institution} {ILL},\ \bibinfo {year}
  {2002})\BibitemShut {NoStop}%
\bibitem [{\citenamefont {Krist}\ \emph {et~al.}(1998)\citenamefont {Krist},
  \citenamefont {Kennedy}, \citenamefont {Hicks},\ and\ \citenamefont
  {Mezei}}]{krist98}%
  \BibitemOpen
  \bibfield  {author} {\bibinfo {author} {\bibfnamefont {T.}~\bibnamefont
  {Krist}}, \bibinfo {author} {\bibfnamefont {S.~J.}\ \bibnamefont {Kennedy}},
  \bibinfo {author} {\bibfnamefont {T.~J.}\ \bibnamefont {Hicks}}, \ and\
  \bibinfo {author} {\bibfnamefont {F.}~\bibnamefont {Mezei}},\ }\href@noop {}
  {\bibfield  {journal} {\bibinfo  {journal} {Phys. B}\ }\textbf {\bibinfo
  {volume} {241}},\ \bibinfo {pages} {82} (\bibinfo {year} {1998})}\BibitemShut
  {NoStop}%
\bibitem [{\citenamefont {Hoghoj}\ \emph {et~al.}(1999)\citenamefont {Hoghoj},
  \citenamefont {Anderson}, \citenamefont {Siebrecht}, \citenamefont {Graf},\
  and\ \citenamefont {Ben-Saidane}}]{hog99}%
  \BibitemOpen
  \bibfield  {author} {\bibinfo {author} {\bibfnamefont {P.}~\bibnamefont
  {Hoghoj}}, \bibinfo {author} {\bibfnamefont {I.}~\bibnamefont {Anderson}},
  \bibinfo {author} {\bibfnamefont {R.}~\bibnamefont {Siebrecht}}, \bibinfo
  {author} {\bibfnamefont {W.}~\bibnamefont {Graf}}, \ and\ \bibinfo {author}
  {\bibfnamefont {K.}~\bibnamefont {Ben-Saidane}},\ }\href@noop {} {\bibfield
  {journal} {\bibinfo  {journal} {Phys. B}\ }\textbf {\bibinfo {volume}
  {267}},\ \bibinfo {pages} {355} (\bibinfo {year} {1999})}\BibitemShut
  {NoStop}%
\bibitem [{\citenamefont {Stunault}\ \emph
  {et~al.}(2006{\natexlab{a}})\citenamefont {Stunault}, \citenamefont
  {Andersen}, \citenamefont {Roux}, \citenamefont {Bigault}, \citenamefont
  {Ben-Saidane},\ and\ \citenamefont {Ronnow}}]{stu06}%
  \BibitemOpen
  \bibfield  {author} {\bibinfo {author} {\bibfnamefont {A.}~\bibnamefont
  {Stunault}}, \bibinfo {author} {\bibfnamefont {K.~H.}\ \bibnamefont
  {Andersen}}, \bibinfo {author} {\bibfnamefont {S.}~\bibnamefont {Roux}},
  \bibinfo {author} {\bibfnamefont {T.}~\bibnamefont {Bigault}}, \bibinfo
  {author} {\bibfnamefont {K.}~\bibnamefont {Ben-Saidane}}, \ and\ \bibinfo
  {author} {\bibfnamefont {H.~M.}\ \bibnamefont {Ronnow}},\ }\href@noop {}
  {\bibfield  {journal} {\bibinfo  {journal} {Phys. B}\ }\textbf {\bibinfo
  {volume} {385}},\ \bibinfo {pages} {1152} (\bibinfo {year}
  {2006}{\natexlab{a}})}\BibitemShut {NoStop}%
\bibitem [{\citenamefont {Bigault}\ \emph {et~al.}(2009)\citenamefont
  {Bigault}, \citenamefont {Andersen}, \citenamefont {Hiess}, \citenamefont
  {Roux}, \citenamefont {Stunault}, \citenamefont {Bisig},\ and\ \citenamefont
  {Bouvier}}]{big09}%
  \BibitemOpen
  \bibfield  {author} {\bibinfo {author} {\bibfnamefont {T.}~\bibnamefont
  {Bigault}}, \bibinfo {author} {\bibfnamefont {K.~H.}\ \bibnamefont
  {Andersen}}, \bibinfo {author} {\bibfnamefont {A.}~\bibnamefont {Hiess}},
  \bibinfo {author} {\bibfnamefont {S.}~\bibnamefont {Roux}}, \bibinfo {author}
  {\bibfnamefont {A.}~\bibnamefont {Stunault}}, \bibinfo {author}
  {\bibfnamefont {G.}~\bibnamefont {Bisig}}, \ and\ \bibinfo {author}
  {\bibfnamefont {A.}~\bibnamefont {Bouvier}},\ }\href@noop {} {\bibfield
  {journal} {\bibinfo  {journal} {ILL Annual Report}\ ,\ \bibinfo {pages} {94}}
  (\bibinfo {year} {2009})}\BibitemShut {NoStop}%
\bibitem [{\citenamefont {Wildes}(2011)}]{wild11}%
  \BibitemOpen
  \bibfield  {author} {\bibinfo {author} {\bibfnamefont {A.~R.}\ \bibnamefont
  {Wildes}},\ }\href@noop {} {\bibfield  {journal} {\bibinfo  {journal}
  {Intern. Tech. Rep. on the performance of the D17 S-bender}\ } (\bibinfo
  {year} {2011})}\BibitemShut {NoStop}%
\bibitem [{\citenamefont {Sears}(1992)}]{sears92}%
  \BibitemOpen
  \bibfield  {author} {\bibinfo {author} {\bibfnamefont {V.~F.}\ \bibnamefont
  {Sears}},\ }\href@noop {} {\bibfield  {journal} {\bibinfo  {journal} {Neutron
  News}\ }\textbf {\bibinfo {volume} {3}},\ \bibinfo {pages} {26} (\bibinfo
  {year} {1992})}\BibitemShut {NoStop}%
\bibitem [{\citenamefont {Stoney}(1909)}]{sto09}%
  \BibitemOpen
  \bibfield  {author} {\bibinfo {author} {\bibfnamefont {G.~G.}\ \bibnamefont
  {Stoney}},\ }\href@noop {} {\bibfield  {journal} {\bibinfo  {journal} {Proc.
  Royal Soc. A}\ }\textbf {\bibinfo {volume} {82}},\ \bibinfo {pages} {172}
  (\bibinfo {year} {1909})}\BibitemShut {NoStop}%
\bibitem [{\citenamefont {Shah}\ \emph {et~al.}(2014)\citenamefont {Shah},
  \citenamefont {Washington}, \citenamefont {Stonaha}, \citenamefont {Ashkar},
  \citenamefont {Kaiser}, \citenamefont {Krist},\ and\ \citenamefont
  {Pynn}}]{shah14}%
  \BibitemOpen
  \bibfield  {author} {\bibinfo {author} {\bibfnamefont {V.~R.}\ \bibnamefont
  {Shah}}, \bibinfo {author} {\bibfnamefont {A.~L.}\ \bibnamefont
  {Washington}}, \bibinfo {author} {\bibfnamefont {P.}~\bibnamefont {Stonaha}},
  \bibinfo {author} {\bibfnamefont {R.}~\bibnamefont {Ashkar}}, \bibinfo
  {author} {\bibfnamefont {H.}~\bibnamefont {Kaiser}}, \bibinfo {author}
  {\bibfnamefont {T.}~\bibnamefont {Krist}}, \ and\ \bibinfo {author}
  {\bibfnamefont {R.}~\bibnamefont {Pynn}},\ }\href@noop {} {\bibfield
  {journal} {\bibinfo  {journal} {Nucl. Instr. Meth. A}\ }\textbf {\bibinfo
  {volume} {768}},\ \bibinfo {pages} {157} (\bibinfo {year}
  {2014})}\BibitemShut {NoStop}%
\bibitem [{\citenamefont {Stunault}\ \emph
  {et~al.}(2006{\natexlab{b}})\citenamefont {Stunault}, \citenamefont
  {Andersen}, \citenamefont {Roux}, \citenamefont {Bigault}, \citenamefont
  {Ben-Saidane},\ and\ \citenamefont {Ronnow}}]{stun06}%
  \BibitemOpen
  \bibfield  {author} {\bibinfo {author} {\bibfnamefont {A.}~\bibnamefont
  {Stunault}}, \bibinfo {author} {\bibfnamefont {K.~H.}\ \bibnamefont
  {Andersen}}, \bibinfo {author} {\bibfnamefont {S.}~\bibnamefont {Roux}},
  \bibinfo {author} {\bibfnamefont {T.}~\bibnamefont {Bigault}}, \bibinfo
  {author} {\bibfnamefont {K.}~\bibnamefont {Ben-Saidane}}, \ and\ \bibinfo
  {author} {\bibfnamefont {H.~M.}\ \bibnamefont {Ronnow}},\ }\href@noop {}
  {\bibfield  {journal} {\bibinfo  {journal} {Phys. B}\ }\textbf {\bibinfo
  {volume} {385}},\ \bibinfo {pages} {1152} (\bibinfo {year}
  {2006}{\natexlab{b}})}\BibitemShut {NoStop}%
\bibitem [{\citenamefont {Moller}, \citenamefont {Passel},\ and\ \citenamefont
  {Stecherr}(1963)}]{mol63}%
  \BibitemOpen
  \bibfield  {author} {\bibinfo {author} {\bibfnamefont {J.~B.}\ \bibnamefont
  {Moller}}, \bibinfo {author} {\bibfnamefont {L.}~\bibnamefont {Passel}}, \
  and\ \bibinfo {author} {\bibfnamefont {F.}~\bibnamefont {Stecherr}},\
  }\href@noop {} {\bibfield  {journal} {\bibinfo  {journal} {J. Nucl. Energy}\
  }\textbf {\bibinfo {volume} {17}},\ \bibinfo {pages} {227} (\bibinfo {year}
  {1963})}\BibitemShut {NoStop}%
\bibitem [{\citenamefont {Scharpf}\ and\ \citenamefont
  {Anderson}(1994)}]{schar94}%
  \BibitemOpen
  \bibfield  {author} {\bibinfo {author} {\bibfnamefont {O.}~\bibnamefont
  {Scharpf}}\ and\ \bibinfo {author} {\bibfnamefont {I.~S.}\ \bibnamefont
  {Anderson}},\ }\href@noop {} {\bibfield  {journal} {\bibinfo  {journal}
  {Phys. B}\ }\textbf {\bibinfo {volume} {198}},\ \bibinfo {pages} {203}
  (\bibinfo {year} {1994})}\BibitemShut {NoStop}%
\bibitem [{\citenamefont {Alianelli}, \citenamefont {del Rio},\ and\
  \citenamefont {Felici}(2004)}]{alia04}%
  \BibitemOpen
  \bibfield  {author} {\bibinfo {author} {\bibfnamefont {L.}~\bibnamefont
  {Alianelli}}, \bibinfo {author} {\bibfnamefont {M.~S.}\ \bibnamefont {del
  Rio}}, \ and\ \bibinfo {author} {\bibfnamefont {R.}~\bibnamefont {Felici}},\
  }\href@noop {} {\bibfield  {journal} {\bibinfo  {journal} {Phys. B}\ }\textbf
  {\bibinfo {volume} {350}},\ \bibinfo {pages} {739} (\bibinfo {year}
  {2004})}\BibitemShut {NoStop}%
\bibitem [{\citenamefont {Windt}(1998)}]{win98}%
  \BibitemOpen
  \bibfield  {author} {\bibinfo {author} {\bibfnamefont {D.~L.}\ \bibnamefont
  {Windt}},\ }\href@noop {} {\bibfield  {journal} {\bibinfo  {journal} {Comput.
  Phys.}\ }\textbf {\bibinfo {volume} {12}},\ \bibinfo {pages} {3} (\bibinfo
  {year} {1998})}\BibitemShut {NoStop}%
\bibitem [{VIT()}]{VITESS}%
  \BibitemOpen
  \href@noop {} {}\Eprint {http://arxiv.org/abs/VITESS website,
  www.hmi.de/projects/ess/vitess} {VITESS website,
  www.hmi.de/projects/ess/vitess} \BibitemShut {NoStop}%
\bibitem [{McS()}]{McStas}%
  \BibitemOpen
  \href@noop {} {}\Eprint {http://arxiv.org/abs/McStas website,
  http://www.mcstas.org} {McStas website, http://www.mcstas.org} \BibitemShut
  {NoStop}%
\bibitem [{\citenamefont {Mishima}, \citenamefont {Ino},\ and\ \citenamefont
  {Sakai}(2009)}]{mish09}%
  \BibitemOpen
  \bibfield  {author} {\bibinfo {author} {\bibfnamefont {K.}~\bibnamefont
  {Mishima}}, \bibinfo {author} {\bibfnamefont {T.}~\bibnamefont {Ino}}, \ and\
  \bibinfo {author} {\bibfnamefont {K.}~\bibnamefont {Sakai}},\ }\href@noop {}
  {\bibfield  {journal} {\bibinfo  {journal} {Nucl. Instr. Meth. A}\ }\textbf
  {\bibinfo {volume} {600}},\ \bibinfo {pages} {342} (\bibinfo {year}
  {2009})}\BibitemShut {NoStop}%
\bibitem [{\citenamefont {Ino}\ \emph {et~al.}(2011)\citenamefont {Ino},
  \citenamefont {Arimoto}, \citenamefont {Yoshioka}, \citenamefont {Mishima},
  \citenamefont {Taketani}, \citenamefont {Muto}, \citenamefont {Shimizu},
  \citenamefont {Kira}, \citenamefont {Sakaguchi}, \citenamefont {Oku},
  \citenamefont {Sakai}, \citenamefont {Shinohara}, \citenamefont {Suzuki},
  \citenamefont {Otono}, \citenamefont {Oide}, \citenamefont {Yamashita},
  \citenamefont {Imajo}, \citenamefont {Funahashi}, \citenamefont {Yamada},
  \citenamefont {Iwashita}, \citenamefont {Kitaguchi}, \citenamefont {Hino},
  \citenamefont {Suzuki}, \citenamefont {Sanuki}, \citenamefont {Seki},
  \citenamefont {Hirota}, \citenamefont {Ikeda}, \citenamefont {Sato},
  \citenamefont {Otake}, \citenamefont {Ohmori}, \citenamefont {Morishima},\
  and\ \citenamefont {Shima}}]{ino11}%
  \BibitemOpen
  \bibfield  {author} {\bibinfo {author} {\bibfnamefont {T.}~\bibnamefont
  {Ino}}, \bibinfo {author} {\bibfnamefont {Y.}~\bibnamefont {Arimoto}},
  \bibinfo {author} {\bibfnamefont {T.}~\bibnamefont {Yoshioka}}, \bibinfo
  {author} {\bibfnamefont {K.}~\bibnamefont {Mishima}}, \bibinfo {author}
  {\bibfnamefont {K.}~\bibnamefont {Taketani}}, \bibinfo {author}
  {\bibfnamefont {S.}~\bibnamefont {Muto}}, \bibinfo {author} {\bibfnamefont
  {H.~M.}\ \bibnamefont {Shimizu}}, \bibinfo {author} {\bibfnamefont
  {H.}~\bibnamefont {Kira}}, \bibinfo {author} {\bibfnamefont {Y.}~\bibnamefont
  {Sakaguchi}}, \bibinfo {author} {\bibfnamefont {T.}~\bibnamefont {Oku}},
  \bibinfo {author} {\bibfnamefont {K.}~\bibnamefont {Sakai}}, \bibinfo
  {author} {\bibfnamefont {T.}~\bibnamefont {Shinohara}}, \bibinfo {author}
  {\bibfnamefont {J.}~\bibnamefont {Suzuki}}, \bibinfo {author} {\bibfnamefont
  {H.}~\bibnamefont {Otono}}, \bibinfo {author} {\bibfnamefont
  {H.}~\bibnamefont {Oide}}, \bibinfo {author} {\bibfnamefont {S.}~\bibnamefont
  {Yamashita}}, \bibinfo {author} {\bibfnamefont {S.}~\bibnamefont {Imajo}},
  \bibinfo {author} {\bibfnamefont {H.}~\bibnamefont {Funahashi}}, \bibinfo
  {author} {\bibfnamefont {M.}~\bibnamefont {Yamada}}, \bibinfo {author}
  {\bibfnamefont {Y.}~\bibnamefont {Iwashita}}, \bibinfo {author}
  {\bibfnamefont {M.}~\bibnamefont {Kitaguchi}}, \bibinfo {author}
  {\bibfnamefont {M.}~\bibnamefont {Hino}}, \bibinfo {author} {\bibfnamefont
  {Z.}~\bibnamefont {Suzuki}}, \bibinfo {author} {\bibfnamefont
  {T.}~\bibnamefont {Sanuki}}, \bibinfo {author} {\bibfnamefont
  {T.}~\bibnamefont {Seki}}, \bibinfo {author} {\bibfnamefont {K.}~\bibnamefont
  {Hirota}}, \bibinfo {author} {\bibfnamefont {K.}~\bibnamefont {Ikeda}},
  \bibinfo {author} {\bibfnamefont {H.}~\bibnamefont {Sato}}, \bibinfo {author}
  {\bibfnamefont {Y.}~\bibnamefont {Otake}}, \bibinfo {author} {\bibfnamefont
  {H.}~\bibnamefont {Ohmori}}, \bibinfo {author} {\bibfnamefont
  {T.}~\bibnamefont {Morishima}}, \ and\ \bibinfo {author} {\bibfnamefont
  {T.}~\bibnamefont {Shima}},\ }\href@noop {} {\bibfield  {journal} {\bibinfo
  {journal} {Phys. B}\ }\textbf {\bibinfo {volume} {406}},\ \bibinfo {pages}
  {2424} (\bibinfo {year} {2011})}\BibitemShut {NoStop}%
\bibitem [{\citenamefont {Pleshanov}(1994)}]{plesh94}%
  \BibitemOpen
  \bibfield  {author} {\bibinfo {author} {\bibfnamefont {N.~K.}\ \bibnamefont
  {Pleshanov}},\ }\href@noop {} {\bibfield  {journal} {\bibinfo  {journal} {Z.
  Phys. B}\ }\textbf {\bibinfo {volume} {94}},\ \bibinfo {pages} {233}
  (\bibinfo {year} {1994})}\BibitemShut {NoStop}%
\bibitem [{\citenamefont {Klauser}\ \emph {et~al.}(2013)\citenamefont
  {Klauser}, \citenamefont {Bigault}, \citenamefont {Rebrova},\ and\
  \citenamefont {Soldner}}]{klau13}%
  \BibitemOpen
  \bibfield  {author} {\bibinfo {author} {\bibfnamefont {C.}~\bibnamefont
  {Klauser}}, \bibinfo {author} {\bibfnamefont {T.}~\bibnamefont {Bigault}},
  \bibinfo {author} {\bibfnamefont {N.}~\bibnamefont {Rebrova}}, \ and\
  \bibinfo {author} {\bibfnamefont {T.}~\bibnamefont {Soldner}},\ }\href@noop
  {} {\bibfield  {journal} {\bibinfo  {journal} {Phys. Proc.}\ }\textbf
  {\bibinfo {volume} {42}},\ \bibinfo {pages} {99} (\bibinfo {year}
  {2013})}\BibitemShut {NoStop}%
\bibitem [{\citenamefont {Klauser}\ \emph {et~al.}(2016)\citenamefont
  {Klauser}, \citenamefont {Bigault}, \citenamefont {Boni}, \citenamefont
  {Coutois}, \citenamefont {Devishvili}, \citenamefont {Rebrova}, \citenamefont
  {Schneider},\ and\ \citenamefont {Soldner}}]{klau16}%
  \BibitemOpen
  \bibfield  {author} {\bibinfo {author} {\bibfnamefont {C.}~\bibnamefont
  {Klauser}}, \bibinfo {author} {\bibfnamefont {T.}~\bibnamefont {Bigault}},
  \bibinfo {author} {\bibfnamefont {P.}~\bibnamefont {Boni}}, \bibinfo {author}
  {\bibfnamefont {P.}~\bibnamefont {Coutois}}, \bibinfo {author} {\bibfnamefont
  {A.}~\bibnamefont {Devishvili}}, \bibinfo {author} {\bibfnamefont
  {N.}~\bibnamefont {Rebrova}}, \bibinfo {author} {\bibfnamefont
  {M.}~\bibnamefont {Schneider}}, \ and\ \bibinfo {author} {\bibfnamefont
  {T.}~\bibnamefont {Soldner}},\ }\href@noop {} {\bibfield  {journal} {\bibinfo
   {journal} {Nucl. Instr. Meth. A}\ }\textbf {\bibinfo {volume} {840}},\
  \bibinfo {pages} {181} (\bibinfo {year} {2016})}\BibitemShut {NoStop}%
\bibitem [{\citenamefont {Halbach}(1980)}]{hal80}%
  \BibitemOpen
  \bibfield  {author} {\bibinfo {author} {\bibfnamefont {K.}~\bibnamefont
  {Halbach}},\ }\href@noop {} {\bibfield  {journal} {\bibinfo  {journal} {Nucl.
  Instr. Meth. A}\ }\textbf {\bibinfo {volume} {169}},\ \bibinfo {pages} {1}
  (\bibinfo {year} {1980})}\BibitemShut {NoStop}%
\bibitem [{rad()}]{radia}%
  \BibitemOpen
  \href@noop {} {}\Eprint
  {http://arxiv.org/abs/http://www.esrf.eu/Accelerators/Groups/InsertionDevices/Software/Radia}
  {http://www.esrf.eu/Accelerators/Groups/InsertionDevices/Software/Radia}
  \BibitemShut {NoStop}%
\bibitem [{\citenamefont {Elleaume}, \citenamefont {Chubar},\ and\
  \citenamefont {Chavanne}(1997)}]{elle97}%
  \BibitemOpen
  \bibfield  {author} {\bibinfo {author} {\bibfnamefont {P.}~\bibnamefont
  {Elleaume}}, \bibinfo {author} {\bibfnamefont {O.}~\bibnamefont {Chubar}}, \
  and\ \bibinfo {author} {\bibfnamefont {J.}~\bibnamefont {Chavanne}},\
  }\href@noop {} {\bibfield  {journal} {\bibinfo  {journal} {Proc. PAC97
  Conf.}\ ,\ \bibinfo {pages} {3509}} (\bibinfo {year} {1997})}\BibitemShut
  {NoStop}%
\end{thebibliography}%

\end{document}